\documentclass[preprint,3p, twocolumn,10pt]{elsarticle}
\usepackage{amssymb}
\usepackage{amsmath}
\usepackage{natbib}
\bibpunct[,]{[}{]}{,}{n}{,}{,}
\usepackage{lineno}

\usepackage{longtable}
\usepackage{pdflscape}
\usepackage{amsmath,amsthm,soul}
\usepackage{amssymb}
\usepackage{graphics}
\usepackage{subfigure}
\usepackage{pifont}
\usepackage{epsfig}
\usepackage{setspace}
\usepackage{graphicx}
\usepackage{enumerate}
\usepackage{multicol}
\usepackage{multirow}
\usepackage{epstopdf}
\usepackage{wrapfig}
\usepackage{float}
\usepackage{caption}
\usepackage{comment}
\usepackage{csquotes}
\usepackage{fixltx2e}
\usepackage{hyperref}
\usepackage{soul}
\usepackage{cancel}
\usepackage{lastpage}

\begin{document}

\begin{frontmatter}



\title{Wind-turbine wake effects on the rate of accumulation of fatigue damage in overhead conductors}

\author[ICL]{Francisco J. G. de Oliveira}
\author[ICL]{Kevin Gouder}
\author[ICL]{Zahra Sharif Khodaei}
\author[ICL]{Oliver R. H. Buxton}

\affiliation[ICL]{organization={Department of Aeronautics},
            addressline={Imperial College London}, 
            city={London},
            postcode={SW7 2BZ}, 
            country={UK}}

\begin{abstract}
Guidance relating to the safe distance within which overhead conductors (OHCs) should not be built with respect to a wind turbine, and vice versa, varies from country to country.
In the United Kingdom the recommendation is that OHCs are not installed within three rotor diameters (3$D$) of wind turbines due to concerns over wake-induced fatigue. 
To assess the physical basis for this recommendation, wind-tunnel experiments were conducted using a scaled wind turbine placed upstream of an instrumented conductor, in conditions representative of forested terrain. 
The conductor was equipped with distributed fibre-optic strain sensing based on Rayleigh backscattering, providing spatially resolved measurements at 2.6 mm resolution. 
Three conductor heights and four turbine-conductor spacings were tested, together with a no-turbine baseline, while maintaining a constant incident wind speed of 7 m\,s$^{-1}$. 
The results identify a critical region near the clamp where mean strain is maximised. 
Neither mean nor fluctuating strain at this location increased substantially due to the presence of the turbine wake. 
Strain fluctuations were dominated by aeolian vibration, with the wake increasing vibration amplitude whilst reducing its characteristic frequency. 
Rainflow counting fatigue analysis shows that damage accumulation depends strongly on conductor height. 
While fatigue damage rates increase when the conductor is fully immersed in the wake, reduced damage rates are observed for lower heights and closer spacings. 
These results suggest that, under forested atmospheric conditions, conductor-turbine separations smaller than the current 3$D$ guidance may be feasible provided the conductor is not fully immersed in the turbine wake.
\end{abstract}


\end{frontmatter}

\section{Introduction} \label{sec:intro}

The adoption of net zero targets for carbon dioxide emissions worldwide will increase the demand for electricity, driven by electrification of sectors such as transport, heating, and industrial processes.
For example, it is estimated that electricity demand in the United Kingdom will increase between 44\% and 150\% by 2050 due to net-zero technological roll outs \citep{barrett2022}.
In addition, the advent of artificial intelligence and the need for energy intensive data centres will further increase this demand with the UK's National Energy System Operator (NESO) estimating that demand from data centres will grow from 5 TW\,hrs annually in 2024 to 22 TW\,hrs by 2030 \citep{NESO2024}.
Whilst these numbers specifically refer to the UK these trends are repeated worldwide.
At the same time that demand for electricity is increasing there is a transition towards renewable energy generation sources.
Wind and solar energy dominate, with the Global Wind Energy Council's 2025 Global Wind Report \citep{GWEC2025} revealing that a record 117 GW of wind power was installed worldwide in 2024 (109 GW onshore) taking the global installed wind capacity to 1,136 GW of which the overwhelming majority (1,052 GW) is installed onshore.
It is thus clear that these pressures will lead to the construction of new overhead transmission lines in proximity to onshore wind turbines, and vice versa, especially in locations with an abundant wind resource such as Scotland.

Since wind turbines harness kinetic energy from the wind they leave a wake downstream, which is a region in which there is a kinetic-energy, and hence momentum deficit in comparison to the unimpeded wind.
The phenomenology of wind-turbine wakes is complex with excellent overviews given by \citet{vermeer2003} and \citet{stevens2017}.
Broadly, they can be divided into two distinct regions, i) the near wake where coherent motions yielded by the blade-tip vortex system, the blade-root vortex system, and the vortex shedding initiated by the nacelle and tower interact with one another \citep[e.g.][]{biswas2024}, and ii) the far wake in which these coherent motions have largely dissipated and the velocity profile of the wake can be approximated through a Gaussian superimposed onto the incoming atmospheric boundary layer (ABL) profile \citep[e.g.][]{bastankhah2014}.
The transition between the near wake and far wake is complex and depends upon the break down of the dominant coherent motions within the wake such as the tip vortex system \citep[e.g.][]{lignarolo2014}. 
It is also dependent on operating conditions, e.g. the turbine's tip-speed ratio \citep{biswas2024}, TSR, $\Lambda = \Omega D / (2U_\infty)$ where $\Omega$ is the rotational speed of the turbine, $D$ is the turbine's diameter, and $U_\infty$ is the incident wind speed, which is the ratio of the solid-body rotational speed of the blade tip to the oncoming wind velocity.
Nevertheless, a good overview of the different phenomenologies of the near wake, transition region, and far wake and where these transitions typically take place is given in \citet{neunaber2020}.
It is important to note that the turbulence level (turbulence intensity $I$) in the incoming flow is an important determinant of where these transitions take place since all wind turbines are situated within the turbulent ABL.

The wake's momentum deficit is at its largest closest to the rotor plane (in the near wake) and then decreases with distance downstream due to the entrainment, and mixing of higher momentum air into the wake \citep[e.g.][]{newman2014, andersen2017,bourhis2025}.
Since the Reynolds number, the ratio between inertial and viscous forces in a fluid flow, is high for these wakes (typically $Re_D \sim \mathcal O(10^7 - 10^8)$ for a utility-scale wind turbine) these wakes are turbulent.
An overhead conductor (OHC) situated within the wake of a wind turbine is thus exposed to a highly unsteady turbulent inflow in which the incident velocity is lower than the unimpeded wind.
Both of these can accelerate the accumulation of fatigue damage in the OHC and lead to premature failure.

Firstly, the unsteady inflow, which varies according to the spectrum of the turbulent motion, yields a fluctuating drag force on the OHC and hence vibration parallel to the wind vector.
This process is known as turbulent buffeting \citep[e.g.][]{weaver1978}.
All OHCs are exposed to turbulent buffeting due to the turbulent nature of the ABL, but the additional wake-turbulence present in a wind-turbine wake can be plausibly expected to modify the buffeting experienced by an OHC situated with an impinging wind-turbine wake.

Secondly, a reduced incident wind velocity (due to an impinging wind-turbine wake) can yield a phenomenon known as aeolian vibration (or alternatively vortex-induced vibration \citep{blevins2001}) .
All OHCs are functionally high aspect ratio cylinders with respect to the incident flow and therefore shed vortices, via a von K\'arm\'an vortex street, which induces an unsteady lift force (perpendicular to the wind vector) on the cable.
The non-dimensional frequency (Strouhal number) at which these vortices are shed, $St_{VS} = f_{VS} d / U_w$ -- where $f_{VS}$ is the vortex shedding frequency, $d$ is the cable diameter, and $U_w$ is the incident wind speed on the cable -- is a function of the Reynolds number of the flow incident on the cable $Re_d$. 
Various aerodynamic regimes, parameterised by $St_{VS}$ and $Re_d$, exist for typical combinations of wind speed and cable diameter yielding vortex shedding in the range $0.17 \lesssim St_{VS} \lesssim 0.21$ \citep{fey1998}.
When this vortex shedding approaches the natural frequency of the cable due to its own inertia/tension then a resonance can be produced that yields (relatively) high-frequency low-amplitude oscillation of the OHC normal to the wind direction.
Accordingly, the variables that affect the nature of the aeolian vibration are the wind speed, outer diameter of the cable, mass per unit length of the cable, and cable-tension.
Aeolian vibration is typically observed for in-service OHCs in a range of moderate wind speeds 0.5--7 m\,s$^{-1}$ \citep{liu2022}.
Accordingly, an OHC immersed in an impinging wind-turbine wake may be pushed into the aeolian vibration regime due to the wake's velocity deficit, even in unimpeded wind speeds that are $> 7$ m\,s$^{-1}$.
Since the wake's velocity deficit varies with distance downstream of the turbine's rotor plane this effect is therefore expected to be a function of the turbine -- conductor spacing.
It has also been noted that increased levels of turbulence intensity have the effect of reducing aeolian vibration on an OHC \citep{wareing2011}, presumably since freestream turbulence is likely to break the spanwise coherence of the vortices shed from the cable thereby minimising the overall fluctuating lift force.
Wind turbine wakes are therefore likely to have two competing effects on an OHC situated within the wake, a likely enhancement due to the wake's velocity deficit and a likely reduction due to the increased turbulence intensity levels in the wake relative to the unimpeded wind.

OHCs exposed to wind turbine wakes may therefore be exposed to i) enhanced turbulent buffeting (due to the wake-added turbulence intensity and modification of the spectrum of turbulent motions with respect to ABL-turbulence), ii) exposure to larger-scale motion caused by the interaction between the OHC and any coherent motions in the turbine wake (e.g. tip vortices) -- this will be a strong function of the spacing between the conductor and the turbine -- and iii) enhanced aeolian vibration.
All three of these mechanisms may accelerate the accumulation of fatigue damage and therefore shorten the in-service lifespan of an OHC placed in close proximity to a wind turbine \citep{wareing2011}.
As a result guidance typically exists that limits the minimum distance that a wind turbine may be built in relation to an overhead power line, and vice versa.
However, the guidance from country to country is inconsistent.
For example, the guidance in the Netherlands makes no reference to aerodynamic effects on the OHC and simply specifies that an OHC must be positioned sufficiently far away from a wind turbine that it will not be affected by mechanical failure of the turbine \citep{RvON2020}.
Conversely, the UK \citep{NG287} specifies a minimum safe distance of three rotor diameters, $3D$, Ireland specifies a minimum safe distance of $3.5D$ \citep{williams2018}, whilst Germany \citep{Germanannexe} specifies a minimum safe distance of $3D$ if the conductors are not damped against wind vibrations and $> 1D$ if the conductors are damped against wind-induced vibrations.
With the expected boom in the building of new onshore wind farms and overhead transmission lines in order to meet net zero targets this guidance can mean having to extend OHC lines by tens of kilometres, at enormous cost, and yet to the authors' knowledge no physical basis has been presented for these minimum safe distances.

Despite this, the literature on the subject is sparse and has been limited to coarse modelling of the effect of a wind turbine wake as a monolithic velocity deficit \citep{williams2018} and some limited field measurements on the Rathrussan -- Shankhill 110 kV line running through the Bindoo wind farm in Ireland \citep{williams2019}.
Both of these studies concluded that the presence of an upstream wind turbine was likely to have a minimal effect on the accumulation of fatigue damage on a nearby OHC, but the uncertainties were very large.
\citet{maciver2014} made some \emph{in situ.} measurements in the Coal Clough wind farm in Northern England and used literature-based analysis to similarly conclude that the effects on fatigue lifetime of an impinging wind-turbine wake would be minimal for typical wood-pole OHCs (i.e. at low height above the ground).

The objective of this work is thus, for the first time, to conduct high-fidelity experiments under controlled laboratory conditions to examine the effect of varying the conductor -- turbine spacing on the accumulation of fatigue damage in an OHC exposed to an impinging wind-turbine wake.
Having conducted these experiments it is hoped that this will be able to provide some physical basis for the guidance issued as to the minimum safe distance between a wind turbine and an OHC, which seems to be converging to $\approx 3D$ from several different countries where aerodynamic effects are considered.
It is further hoped that the relative importance of aeolian vibration, which could be either enhanced by the turbine-wake's elevated velocity deficit or reduced by the elevated levels of wake turbulence, and turbulent buffeting (expected to be increased by the incresed wake-turbulence levels) could be explored with this data set.
This manuscript sets out the design and conduct of these experiments and some relative fatigue analysis.

\section{Scope of works}

The objective was to simulate, as closely as possible, the Gordon Bush wind farm in Scotland, requiring the generation of turbulent conditions within the wind tunnel which were reminiscent of open moorland and forested areas.
The experiments called for a single wind turbine model to be placed upstream of a representative overhead conductor.
The real-life wind turbines have the following full-scale properties
\begin{itemize}
\item Rotor diameter: $D=$ 155 m
\item Hub height: $h =$ 122.5 m
\item Maximum blade-tip height: 200 m
\item Minimum blade-tip separation to the ground: 45 m
\end{itemize}

The OHC was to be representative of either i) 700 mm$^2$ all aluminium-alloy conductor (AAAC) ARAUCARIA and ii) 300 mm$^2$ AAAC UPAS with a span of $L =$ 350 m.
Three different heights above the ground for the conductor were to be considered, $H \in \{30 \text{ m}, 58 \text{ m}, 113 \text{ m}\}$. 
Both the lift and the drag on the cable scale with $U_w^2$ and so it was presumed that higher wind speeds would yield a greater response from the cable, both the turbulent buffeting and the aeolian vibration.
The wind speed was thus selected to be $U_w = 7$ m\,s$^{-1}$, which is towards the upper end of the band of wind speeds thought to be most responsible for aeolian vibration.
Importantly, this was the wind speed incident on the cable, not the wind speed at the hub height of the turbine.

The important free parameter in the test campaign was the separation between the conductor and the rotor plane of the wind turbine.
For each conductor height five different experiments were planned.
The first of which simply exposed the conductor to the atmospheric-like flow in the absence of a wind turbine.
This case is to be used as the baseline.
Four turbine -- conductor separations were then to be tested with the separation distance $x_C \in \{1.5D, 2D, 3D, 4D \}$, where $D$ is the rotor diameter, i.e. distances that correspond to $x_C \in \{183.75 \text{ m}, 245 \text{ m}, 367.5 \text{ m}, 490 \text{ m}\}$ between the OHC and the turbine.

\section{Measurement techniques} \label{sec:meas}

The experimental campaign made use of Rayleigh backscattering fibre optic sensing (RBS) in order to measure the fluctuating distributed strain along the span of the cable.
A fibre optic was bonded to the conductor and a monochromatic light source is shone down the fibre.
These sensors leverage a fundamental optical phenomenon that occurs throughout the fibre optic medium. 
Rayleigh backscattering arises from the interaction between light waves and microscopic density fluctuations in the transmitting material, i.e. the fibre. 
As the light beam propagates within the fibre, it encounters density variations, that create local scattering points within the fibre. 
The Rayleigh backscattered photons originating from these scattering regions travel back along the incident beam's path. 
The frequency of the backscattered beams is influenced by mechanical stress applied to the fibre (i.e. those induced by the velocity field adjacent to the conductor/fibre), which in turn affects the fibre’s density fluctuations. 

An optical frequency-domain reflectometry (OFDR) system exploits the Rayleigh backscattering spectrum occurring in a single-mode fibre (SMF), generated by the
random modulation of the refractive index profile along the length of the fibre, to instrument and capture details of the structural response within the fibre \citep{kreger,kwon19}.
The OFDR system splits the light beam into two portions with the same spectral component but different amplitudes using an optical coupler. 
The beam with the larger amplitude is directed to the fibre sensing path, while the other serves as a reference signal. 
A beat signal is employed to identify different positions along the SMF \citep{li23}. 
As various positions of the fibre are exposed to external loads, Rayleigh backscattered light experiences the Rayleigh frequency shift of the beat signal for each position, correlating with locally applied strain using a pre-determined strain-frequency shift of the fibre \citep{li23,xu2020,Berrocal02012021,li_real_time_2026}. 
The portions of the fibre bonded to the conductor capture the structural response with fine spatial and temporal resolution, enabling the extraction of flow-induced effects along the fibre path direction. 
The frequency of acquisition of the RBS is limited by the OFDR system, currently set at 100 Hz for 20 m length mode fibres in the equipment used in the laboratory.
Despite the moderate frequency of acquisition, the sensor network enables the structural response to be captured in a time-resolved fashion with a fine spatial resolution through a single channel.
In the current set of experiments the conductor was sampled at 432 locations (see \S\ref{sec:instrument}).
Achieving concurrent measurements at as many locations (with such a fine spatial resolution) using alternative instrumentation, e.g. strain gauges (or even fibre Bragg grating sensors, FBGs \cite{zhou1999,zhou2000}) is intractable without severely compromising the inertia of the cable, and hence its dynamical response to cross flow.
More details on the pioneering use of RBS coupled to flow measurements for use with bodies exposed to turbulent inflows can be found in our previous works \citep{oliveira24, oliveira25}.

\section{Scaling the experiments down to the wind tunnel}

The wind tunnel used was the $10' \times 5'$ wind tunnel hosted within the Aeronautics Department at Imperial College London.
The advantages of this facility are:
\begin{enumerate}
\renewcommand{\labelenumi}{\roman{enumi}. }
\item the large test section, meaning that large models of wind turbines can be used without blockage effects. 
A large turbine model means that we can use a large model for the conductor whilst preserving the ratio of $L/D$.
A larger model will mean larger strains which will increase the signal to noise ratio for our RBS measurements,
\item the long fetch of the test section meaning that various surface-roughness elements and spires can be used to faithfully recreate atmospheric-like turbulence,
\item it is temperature controlled (via a 200 kW heat exchanger) that ensures the operating temperature remains constant which is important to ensure that any signal detected by our fiber-based instrumentation is due to imposed strains as opposed to temperature fluctuations.
\end{enumerate}

\subsection{Scaling of the turbine model and conductor span/height}
\begin{figure*}   
\begin{center}
\includegraphics[width=\linewidth]{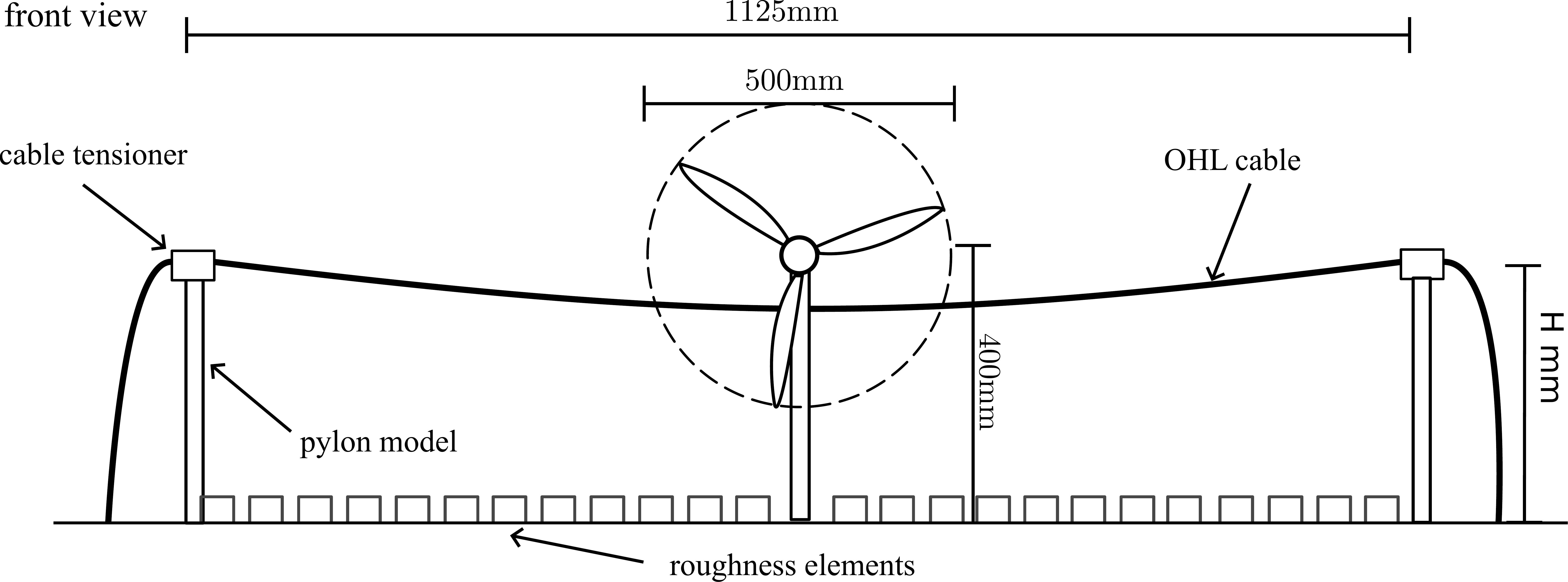}
\caption{Front on schematic of the experimental setup in the wind tunnel.}
\label{fig:front}
\end{center}
\end{figure*}

The cross section of the wind tunnel test section is 3.04 m $\times$ 1.52 m.
Wind tunnel testing requires that the blockage of the tunnel must be typically kept below 5\% which necessitated the use of a model wind turbine with turbine diameter $D = 500$ mm.
In order to preserve the ratios of hub height to turbine diameter ($h/D$) the hub height was set to $h = 400$ mm, and conductor span to turbine diameter ($L/D$) the span was set to $L = 1.125$ m.
The schematic for this can be observed in figure \ref{fig:front}.

\begin{figure}[h]
\begin{center}
\includegraphics[width=\linewidth]{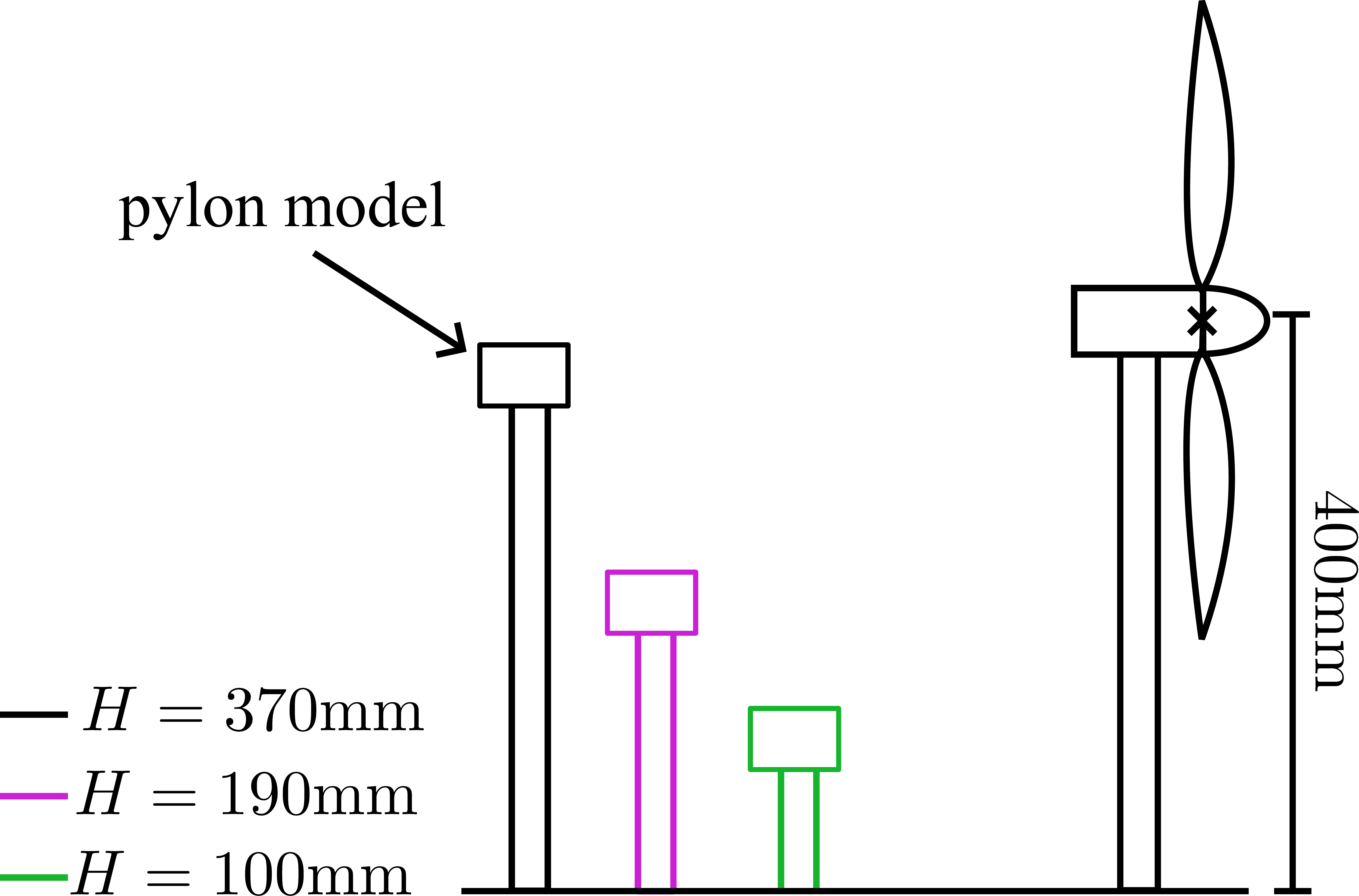}
\caption{Side on schematic of the experimental setup in the wind tunnel giving an indication of the conductor heights $H$ tested. Note that the separation between the turbine and the conductors is not to scale.}
\label{fig:conheight}
\end{center}
\end{figure}

It was likewise necessary to preserve the ratios for the heights of the conductor above the ground to the turbine diameter ($H^\star = H/D$).
For the three different conductor heights (30 m, 58 m, 113 m) the conductor was thus mounted at $H \in$ \{100 mm, 190 mm, 370 mm\}, illustrated in figure \ref{fig:conheight}.
Note that the conductor height is measured to the height of the pylons.
There will be some sag in the cable (which is discussed in \S\ref{sec:condscal}) and so the centre of the cable will always be below height $H$.

\begin{figure}[h]
\begin{center}
\includegraphics[width=\linewidth]{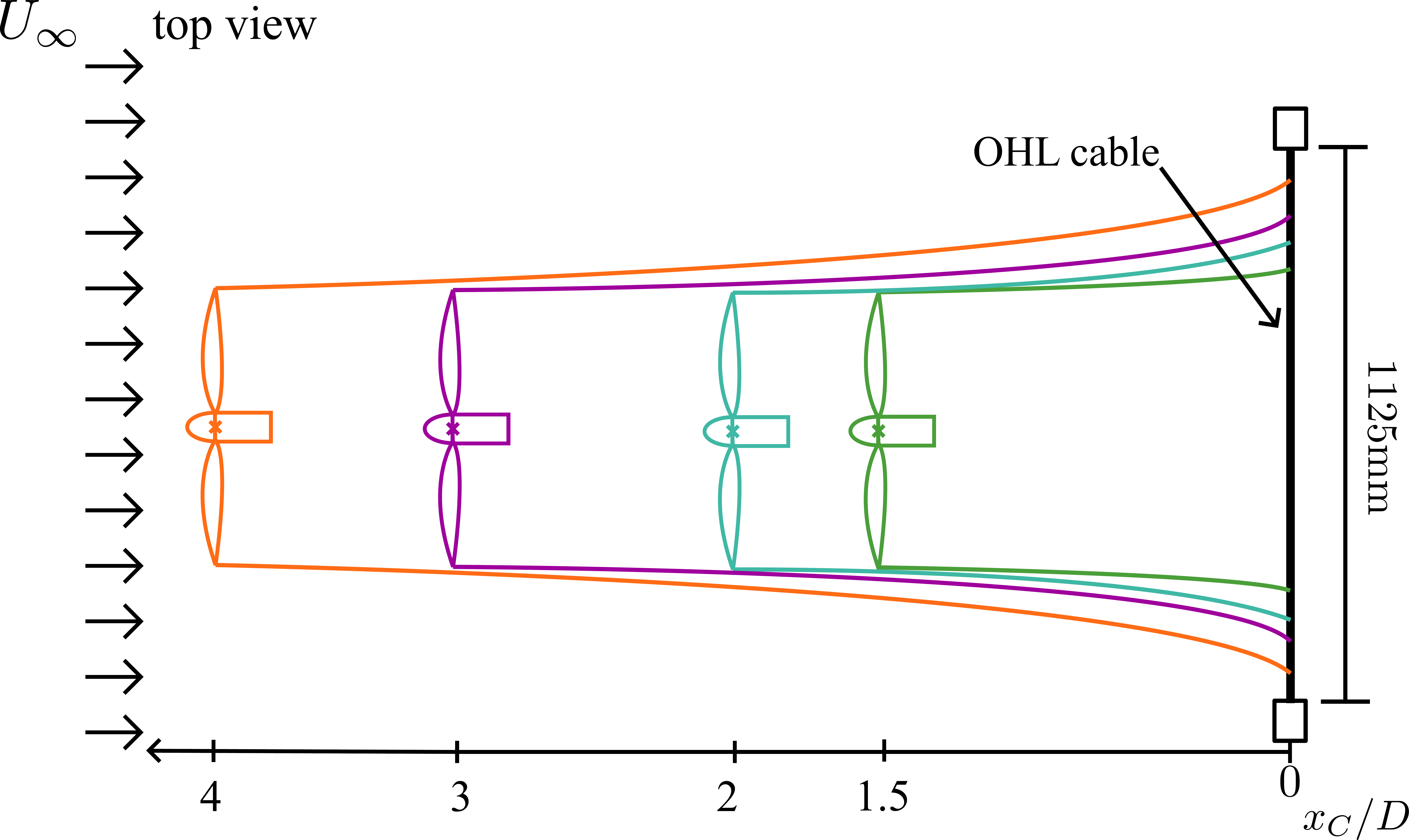}
\caption{Top view schematic of the experimental setup in the wind tunnel giving an indication of the spreading of the wind-turbine wake, and the proportion of the conductor span that is exposed to the wake. 
Note that at the lowest conductor height it is expected that the wake will pass over the conductor for the two smallest turbine -- conductor spacings.
Not to scale.}
\label{fig:wake}
\end{center}
\end{figure}

The separation between the conductor and the plane of the rotor is also preserved with respect to the turbine diameter, i.e. distances of $\{1.5D, 2D, 3D, 4D \}$ correspond to separations of \{0.75 m, 1m, 1.5 m, 2 m\}.
It is important to preserve this ratio since the width (and spreading) of the turbine wake is expected to scale with the turbine diameter, meaning that the proportion of the conductor span that is subjected to the wake is determined by the turbine diameter and the streamwise distance between the rotor plane and the conductor $x_C$.
This is illustrated in figure \ref{fig:wake}.

\subsection{Simulation of atmospheric-like turbulent winds}
The team working in the $10' \times 5'$ tunnel have extensive experience in generating atmospheric-like turbulence to simulate flows over various terrains, such as urban, forested, coastal, and offshore \citep{gouder24}.
There are various ways to categorise the turbulence in lightly forested areas, which is characteristic of the terrain in proximity to the Gordon Bush wind farm.
ESDU data item ESDU 82026 \citep{esdu82026} specifies a surface roughness of $\approx 0.2$ m for regions with extensive tree coverage.
An alternative specification is via the Hellmann exponent which specifies the wind shear in a boundary layer by considering the difference between the mean velocities, $V_1$ and $V_2$ at two different heights ($z_1$ and $z_2$) above the ground, i.e. 
\begin{equation}
\frac{V_2}{V_1} = \left (\frac{z_2}{z_1} \right )^ \alpha
\end{equation}
where $\alpha$ is the Hellmann exponent.
Forested regions are expected to have a Hellmann exponent in the range $0.2 \lesssim \alpha \lesssim 0.25$ \citep{banuelos2011}.

\begin{figure}[h]
\begin{center}
\includegraphics[width=\linewidth]{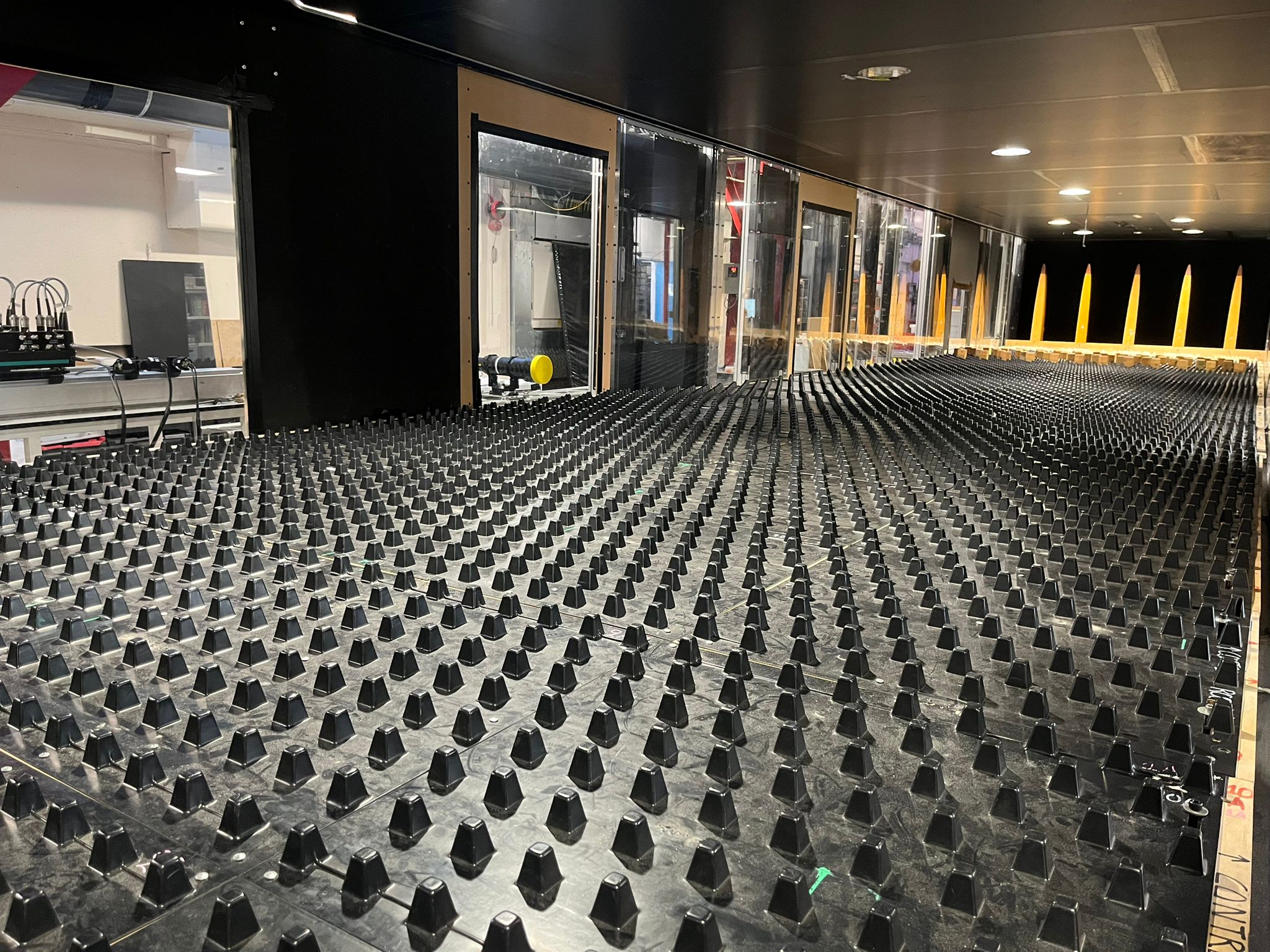}
\caption{Photograph of the surface roughness used in the wind tunnel to simulate an atmospheric flow over forested terrain. Visible far upstream, at the inlet to the test section, are the 1 m high Counihan-type spires and the fence. The cube roughness is visible covering the majority of the floor of the test section.}
\label{fig:roughness}
\end{center}
\end{figure}

\begin{figure}[ht]
\begin{center}
\includegraphics[width=\linewidth]{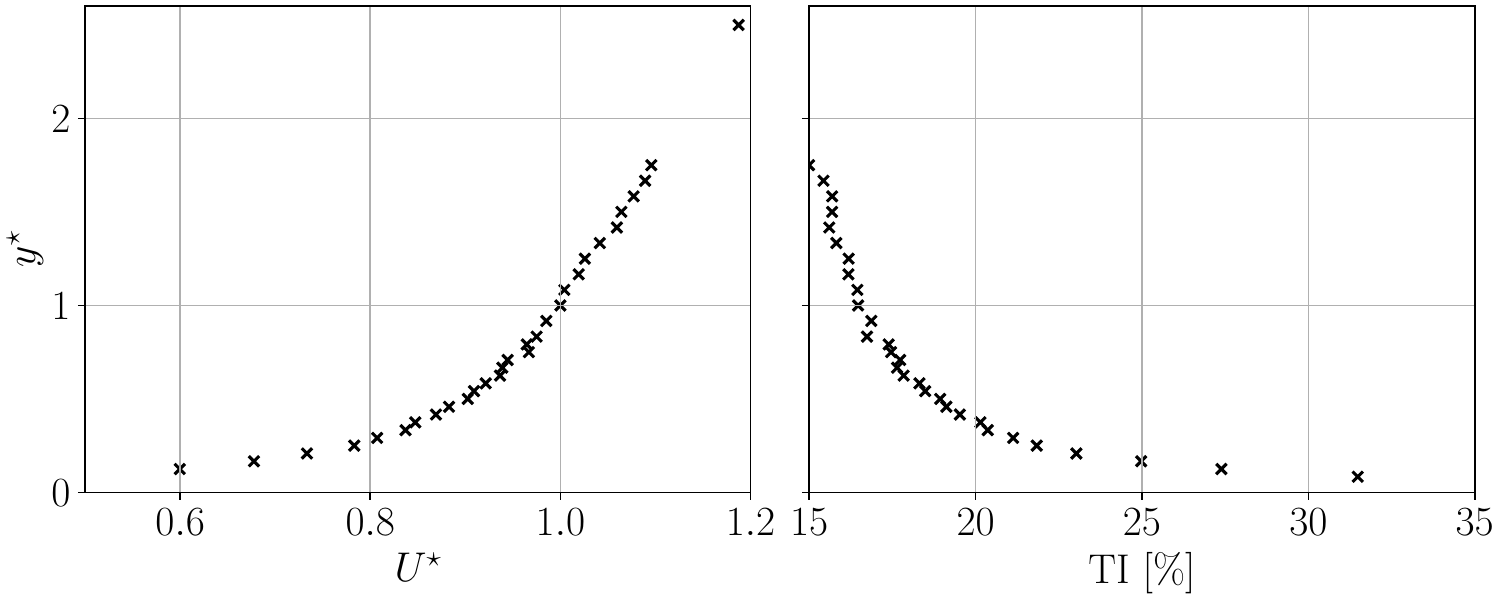}
\caption{(a) Non-dimensional velocity profile $U^\star$ as a function of non-dimensional distance from the ground $y^\star$. $y^\star = 1$ corresponds to the hub height of the turbine (i.e. $y = h$, simulating 122.5 m above the ground) with the velocity $U$ normalised by the mean wind speed at the hub height, i.e. $U^\star(y^\star=1) = 1$. (b) Turbulence intensity profile in the boundary layer as a function of $y^\star$.}
\label{fig:velprof}
\end{center}
\end{figure}

In order to simulate a representative atmospheric flow characteristic of forested terrain a combination of spires at the entry to the wind-tunnel test section, a fence immediately downstream of the spires, and surface-mounted cube roughness were used.
This configuration is visible in figure \ref{fig:roughness}.
The spires were of the Counihan type \citep{counihan1969} and of height 1 m, the fence had a height of 37.5 cm, and the roughness blocks had sizes 32 mm $\times$ 32 mm $\times$ 32 mm.
The developed mean velocity profile for the boundary layer created with this setup is presented in the left-hand pane of figure \ref{fig:velprof} and the profile of turbulence intensity is presented in the right-hand pane.
The computed Hellmann exponent was $\alpha = 0.22$ which lies in the middle of the expected range for flows over forested terrain.

\subsection{Scaling of the conductor diameter and wind speed selection} \label{sec:condscal}

It is not realistic, practical, or desirable to retain the same ratio of conductor diameter to turbine diameter $d/D$ in the scaled-down wind-tunnel experiments.
For example, were we to preserve this diameter ratio for a UPAS conductor then the scaled-down conductor would have a diameter in the wind tunnel of only 93.5 $\mu$m, comparable to the thickness of a human hair.
The two phenomena that we are interested in in this experimental campaign are aeolian vibration and turbulent buffeting.
The former is caused by a resonance between the aerodynamic forcing frequency of the cable, due to the unsteady vortex street within the wake that the cable produces, and the cable's natural frequency.
Cables only produce unsteady vortex streets in their wake if they are in an aerodynamic regime that yields such flow unsteadiness.
A 93.5 $\mu$m cable would be at too low a Reynolds number, and therefore in an aerodynamic regime that does not yield vortex shedding and hence no aeolian vibration.
This necessitates using a cable with a larger diameter $d$ and thus larger ratio $d/D$.

The range of wind speeds that are typically considered to be responsible for aeolian vibration are between 1 and 7 m\,s$^{-1}$.
Accordingly, one of the mechanisms through which wind turbines are able to induce aeolian vibration in conductors situated in their wake is through reducing the incident wind speed on the conductor into this range by virtue of the velocity deficit in the turbine wake.
This means that for better dynamic matching in the wind tunnel it is desirable to maintain the ratio of the aerodynamic forcing frequency on the cable, $f_{VS}$ (for vortex shedding frequency) to the natural frequency $f_{NAT}$.

The natural frequency of the cable is determined as follows 
\begin{equation}
f_{NAT} = \frac{N}{2L} \sqrt{\frac{T_{eff}}{\mu}}
\end{equation} 
where $T_{eff}$ is the effective tension applied to the cable at its two terminations, $\mu$ is the mass of the cable per unit length, $L$ is the span of the cable, and $N$ is the vibration mode.
Relevant information for the real-life conductors, provided in the schedule of works, is presented in table \ref{tab:cables}.

\begin{table*}[t]
\begin{center}
\begin{tabular}{l|c|c|c|c|c|c} 
\multirow{2}{*}{\textbf{Cable type}} & \multirow{2}{*}{$\mu$ (kg m$^{-1}$)} & \multirow{2}{*}{$T$ (kN)} & \multirow{2}{*}{$d$ (mm)} & \multirow{2}{*}{$f_{NAT}$ (Hz)} & \multicolumn{2}{c}{$f_{VS}$ (Hz)} \\ \cline{6-7}
& & & & & @ 7 m s$^{-1}$ & @ 15 m s$^{-1}$ \\ \hline \hline  
UPAS & 1.021 & 29.39 & 28.98 & 0.2424$N$ & 46.7 & 100.1 \\ \hline
ARAUCARIA & 2.345 & 48.45 & 37.3 & 0.2053$N$ & 37.5 & 80.43
\end{tabular}
\caption{Conductor properties.}
\label{tab:cables}
\end{center}
\end{table*}

The last column of table \ref{tab:cables} gives us the aerodynamic forcing frequency due to the vortex shedding from the cable.
From an aerodynamic perspective, we can model the cable as a cylinder.
The vortex shedding frequency for a cylinder exposed to a uniform cross-flow is typically described in terms of its Strouhal number,
\begin{equation}
St_{VS} = \frac{d f_{VS}}{U_w}.
\end{equation}
$St_{VS}$ is a function of the Reynolds number of flow past the cylinder
\begin{equation}
Re = \frac{U_w d}{\nu}
\end{equation}
recalling that $U_w$ is the wind speed incident on the cable, $d$ is the cable diameter, and $\nu$ is the kinematic viscosity.
At an incident wind speed of $U_w = 7$ m s$^{-1}$, the Reynolds number for the two conductors lies in the range of 14,000 to 18,000.
In this Reynolds number range the expected vortex shedding Strouhal number is $St_{VS} \approx 0.19$ \citep{fey1998}.
Knowledge of this Strouhal number, and the assumed wind speed (7 m\,s$^{-1}$) and cable diameter enables us to compute the vortex shedding frequency for the two different conductors in the two right-hand columns of table \ref{tab:cables}.

Assuming that $N=1$ is the most energetic vibration mode for the cable then we can compute the ratio of the vortex shedding frequency (which is the frequency at which the conductor is aerodynamically forced) to the natural frequency (due to the cable's inertia) 
\begin{equation}
R_F = \frac{f_{VS}}{f_{NAT}} [@ U_\infty = 7 \text{m\,s}^{-1} ] \approx 190. \label{eq:rf}
\end{equation}
The wind tunnel experiments thus aim to match this ratio to ensure dynamical similarity.

For a given cable span (fixed by the scope of the works to be 1.125 m in the wind tunnel, matching the ratio of $L/D$ for the real span of $L = 350$ m and $D = 155$ m) we may fix the cable's natural frequency by varying $\mu$ (effectively the material out of which the cable is made), $d$, and $T$.
We wish for the model conductor to be in a similar aerodynamic regime (parameterised by the Reynolds number) since this will yield similar unsteady vortices in the wake of the model conductor to what is found in the wake of the real conductor.
This motivates the use of a cable with diameter $d = 2.62$ mm, which yields Kelvin-Helmholtz vortices in the shear layers separating from the cable.
Whilst this is a lower Reynolds number to that for the real conductor, and is in a slightly different aerodynamic regime, it is nevertheless similar with $St_{VS} \approx 0.21$ for the model as opposed to $St_{VS} \approx 0.19$ for the real conductor.
A further advantage of selecting a sufficiently large $d$ is that bonding the fibre optic to the cable, as is necessary for the RBS measurements, would not have fundamentally altered its aerodynamic shape, in particular the symmetry of the cable since asymmetries lead to mean aerodynamic loads (lift forces).
The ratio of the fibre diameter to the cable diameter is also sufficiently small to ensure that the fibre does not act as an aerodynamic trip and therefore ``artificially'' push the cable into a different aerodynamic regime.

A cable made of ethylene propylene diene monomer (EPDM) rubber was selected, fixing $\mu$.
The advantages of this material are that it is i) smooth meaning that bonding the fibre optic to it is straightforward, and ii) being made of a polymer it will enhance the signal to noise ratio of the acquired strain measurements due to a relatively low Young's modulus.
For the given diameter $d = 2.62$ mm, this cable has a mass per unit length of $\mu = 1.014 \times 10^{-2}$ kg\,m$^{-1}$.
This most closely resembled the ARAUCARIA conductor, although it should be pointed out that the flow physics for both conductors are similar, with only a small difference in the ratio $R_F$ between the two ($R_F = 182$ for ARAUCARIA and $R_F = 192$ for UPAS).

Preserving the ratio $R_F$ between the real wind turbine and the wind-tunnel model turbine thus hinges on setting the correct tension in the cable, $T$.
Rearranging our definition of the vortex shedding Strouhal number gives us the vortex shedding frequency
\begin{equation}
f_{VS} = \frac{St_{VS} U_\infty}{d}.
\end{equation}
We can now define the natural frequency of the cable as a function of $\{f_{VS}, U_\infty, d \}$
\begin{equation}
f_{NAT} = \frac{St_{VS} U_\infty}{d R_F}. \label{eq:fnat}
\end{equation}
The tension is defined as a function of $f_{NAT}$,
\begin{equation}
T = 4L^2 \mu f_{NAT}^2
\end{equation}
assuming that the most energetic vibration mode is $N = 1$.
In combination with \eqref{eq:fnat}, and the definition for $R_F$ given in \eqref{eq:rf}, the tension can be re-written as
\begin{equation}
T = \frac{4L^2 \mu St_{VS}^2 U_\infty^2}{d^2 R_F^2}. \label{eq:tens}
\end{equation}
From a practical perspective, the tension in the cable will manifest as the maximum sag of the cable (at the centre), which is given by
\begin{equation}
y_{sag} = \frac{\mu g L^2}{8T} .
\end{equation}
We may now input the definition of the tension $T$ that maintains the ratio $R_F$ into this definition for the sag
\begin{align}
y_{sag} &= \frac{gd^2 R_F^2}{32 St_{VS}^2 U_\infty^2} = \frac{g}{32 f_{NAT}^2} \label{eq:sag} \\ 
\therefore f_{NAT} &= \frac{f_{VS}}{R_F} = \sqrt{\frac{g}{32 y_{sag}}}.
\end{align}
\eqref{eq:sag} tells us that if we select the diameter of the cable and the wind speed in the tunnel (and in selecting these two parameters we in turn set $St_{VS}$) then we can ensure that the ratio $R_F$ between the aerodynamic forcing frequency and the natural frequency of the cable due to its inertia is set by adjusting the tension in the cable to give us the correct sag in the middle of the cable $y_{sag}$.
An alternative way of looking at this is that having selected the size and material of the cable then we can adjust the sag to give us the correct natural frequency that preserves ratio $R_F$

\section{Experimental methods}

\subsection{Design of the wind turbine model}

\begin{figure}[h]
\begin{center}
\includegraphics[width=\linewidth]{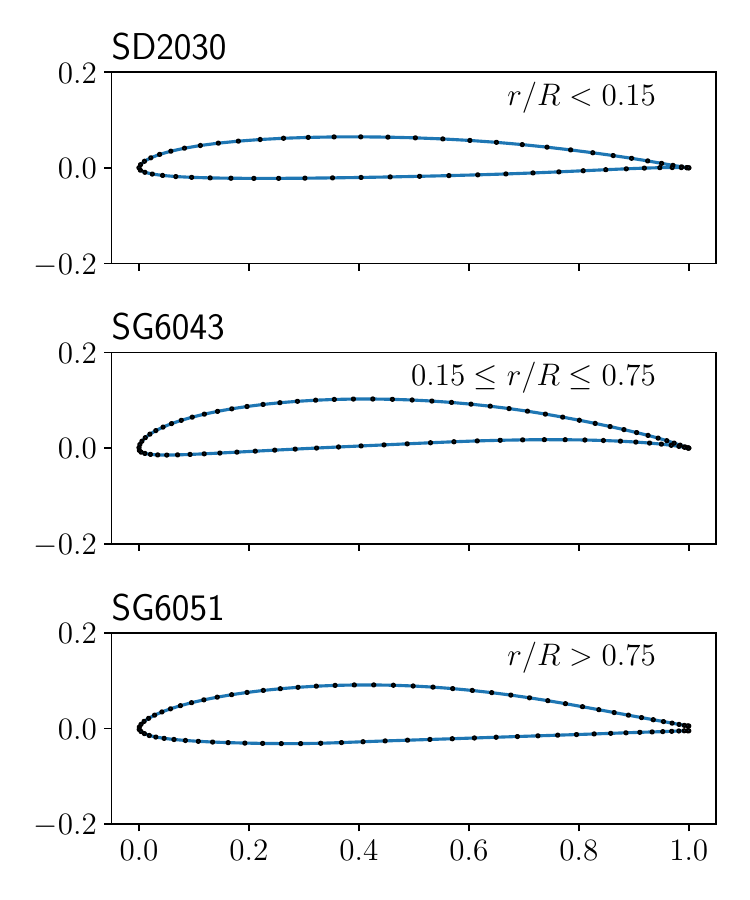}
\caption{The aerofoil sections used in the design of the wind-turbine model. Top: the blade section close to the hub, middle: the middle section of the blade, bottom: the blade section close to the tip.}
\label{fig:aerofoils}
\end{center}
\end{figure}

Since the wind turbine model has been scaled down to fit into the wind tunnel the operating Reynolds number has been significantly reduced, from $\mathcal O (10^8)$ to $\mathcal O(10^5)$.
Aerofoil sections are designed to operate in particular aerodynamic regimes, parameterised by the Reynolds number, and so exact replicas of the aerofoils used in a real wind turbine (operating in a Reynolds number range that is orders of magnitude greater) would not be efficient in the wind-tunnel setting.
Accordingly, three different aerofoil sections, that are efficient at lower Reynolds numbers, were used in the design of the turbine.
These aerofoil sections are illustrated in figure \ref{fig:aerofoils}.

An in-house blade element momentum (BEM) code was used for the aerodynamic design of the turbine.
The turbine was designed to be optimised to a tip speed ratio $\Lambda = 4$.
The TSR of the turbine, or more specifically $\Omega$ given a choice of $U_\infty$, was controlled by using an electric motor running as a generator that is built into the turbine model to apply a braking torque to the shaft.
When operating at the optimum TSR the turbine extracts the most power from the flow, and hence this is the preferred operating condition for wind turbines in wind speeds that are below the rated wind speed.
The scope of works required simulating wind speeds yielding  $U_w = 7$ m\,s$^{-1}$ incident on the conductor which is well below the rated wind speed for typical MW-rated utility wind turbines.
Accordingly, in these conditions the turbines would be operating at the optimum tip speed ratio, which is thereby recreated in the experimental campaign.

When varying the separation between the turbine and the conductor it is more convenient to leave the conductor fixed in place, since this is instrumented with the measurement equipment, and to move the turbine with respect to the fixed location of the conductor.
The correct turbine -- conductor separations were measured out and the turbine model was moved, and bolted down into the correct position between experimental runs.

\subsection{Design of the conductor -- pylons assembly}

\begin{figure}[h]
\begin{center}
\subfigure[]{\includegraphics[width=0.472\linewidth]{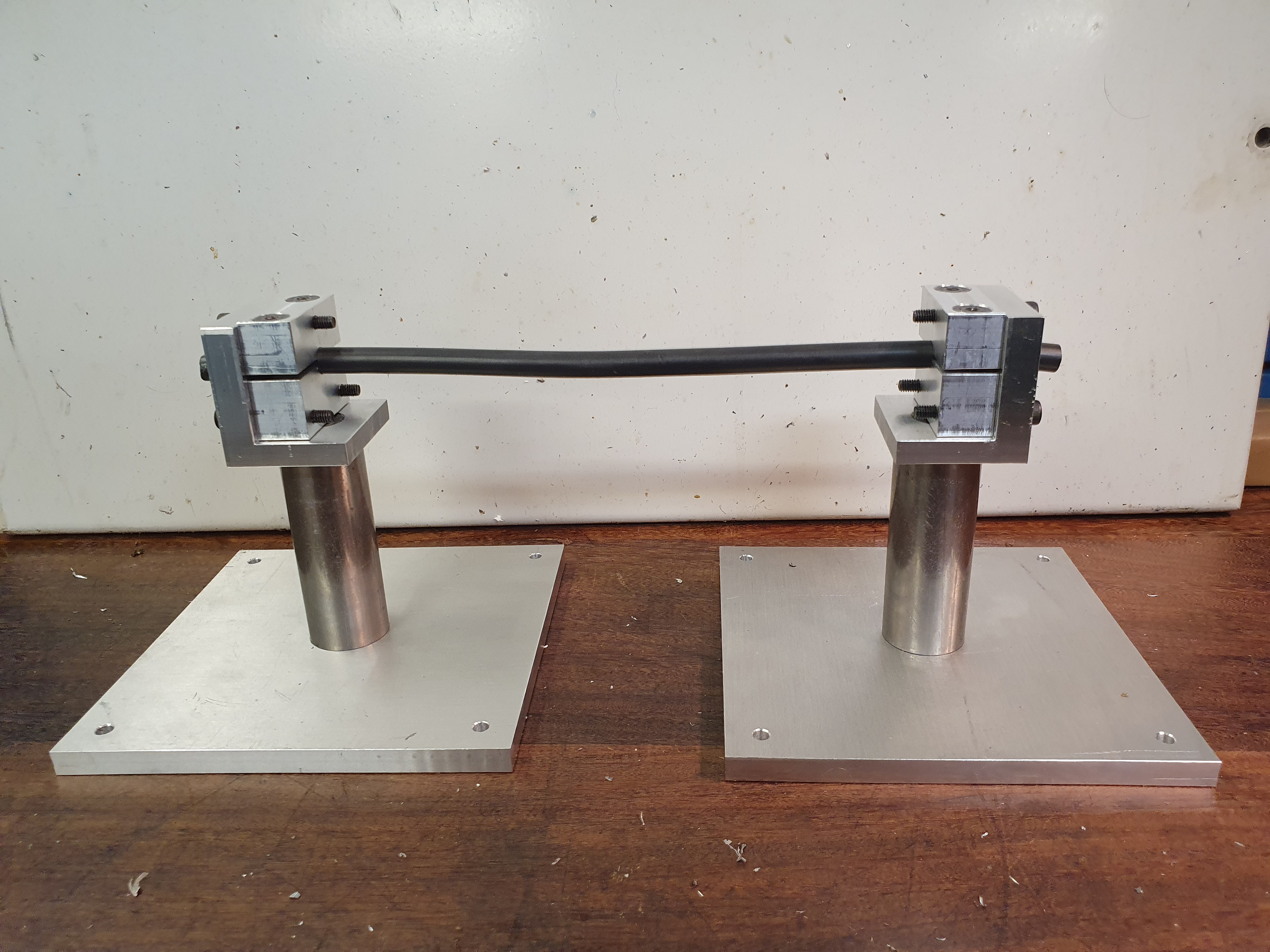}}
\subfigure[]{\includegraphics[width=0.4\linewidth]{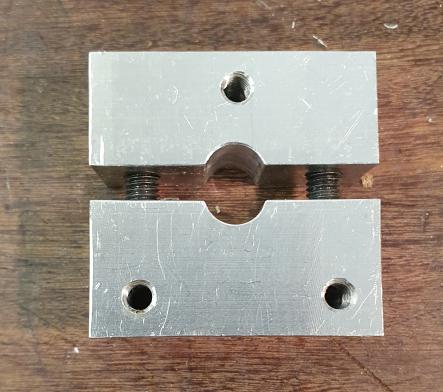}}
\caption{(a) Side view of the simple pylon assembly. Note that the cable span is much smaller, and the diameter larger, than that used in the experimental campaign. (b) Close up of the clamping mechanism that was used to set the tension in the cable.}
\label{fig:pylon}
\end{center}
\end{figure}

The focus of this work is not on the effect of the turbine wake on the pylons, and so a scaled model of a real pylon was not necessary.
A simple configuration, illustrated in figure \ref{fig:pylon}, was used in which a purpose-built clamping mechanism was placed atop a simple cylindrical metal stand.
As discussed in \S\ref{sec:condscal} the correct tension was set in the cable by adjusting the sag at the middle of the cable to ensure that ratio $R_F$ is preserved between the wind-tunnel model and the real-life wind turbine.
The procedure for setting the correct tension in the conductor model was as follows:
\begin{enumerate}
\item Both ends of the cable were clamped in the clamping mechanism.
The clamping force were not specified precisely, other than to ensure that the cable was not deformed plastically or broken.
\item The screws in the clamping mechanism were adjusted in order to modify the sag (tension) in the cable.
\item The correct sag (and therefore tension) was set by exciting the first vibrational mode of the cable and the natural frequency measured by examining the output from the RBS sensors.
\item This process was iterated until the cable had the correct natural frequency $f_{NAT}$ and thus the correct tension \eqref{eq:tens}, or alternatively sag \eqref{eq:sag}.
\end{enumerate}

\subsection{Instrumentation of the conductor model} \label{sec:instrument}

The RBS sensors were embedded onto the cable's surface.
The single-mode fibre (SMF) used in this set of experiments was a SM1500 (9/125)P, by FIBRECORE, possessing a polyimide coating with bend insensitivity and enhanced photosensitivity, with a diameter of 125 $\mu$m. 
The fibre was bonded to the cable with cyanoacrylate glue so that the entire span of the conductor was instrumented (i.e. over a length of 1.125 m) with the fibre sampled at distances of $\Delta y = 2.6$ mm, giving a total of 432 data points along the span of the conductor, regardless of conductor height $H$.
The small diameter and weight of the fibre did not affect the aerodynamics of the cable, nor its natural frequency due to modifications to its inertia.
The wind tunnel was temperature controlled meaning that  strain measurements obtained by the system consist solely of axial strains within the cable induced by the flow. 
The strain data were acquired with the OFDR system developed by LUNA Ltd (ODiSI-B) and was sampled at a frequency of \mbox{$f_{acq.} = $ 100 Hz}, giving us full temporal resolution for dynamics slower than 50 Hz. 

\begin{figure}[h]
\begin{center}
\includegraphics[width=\linewidth]{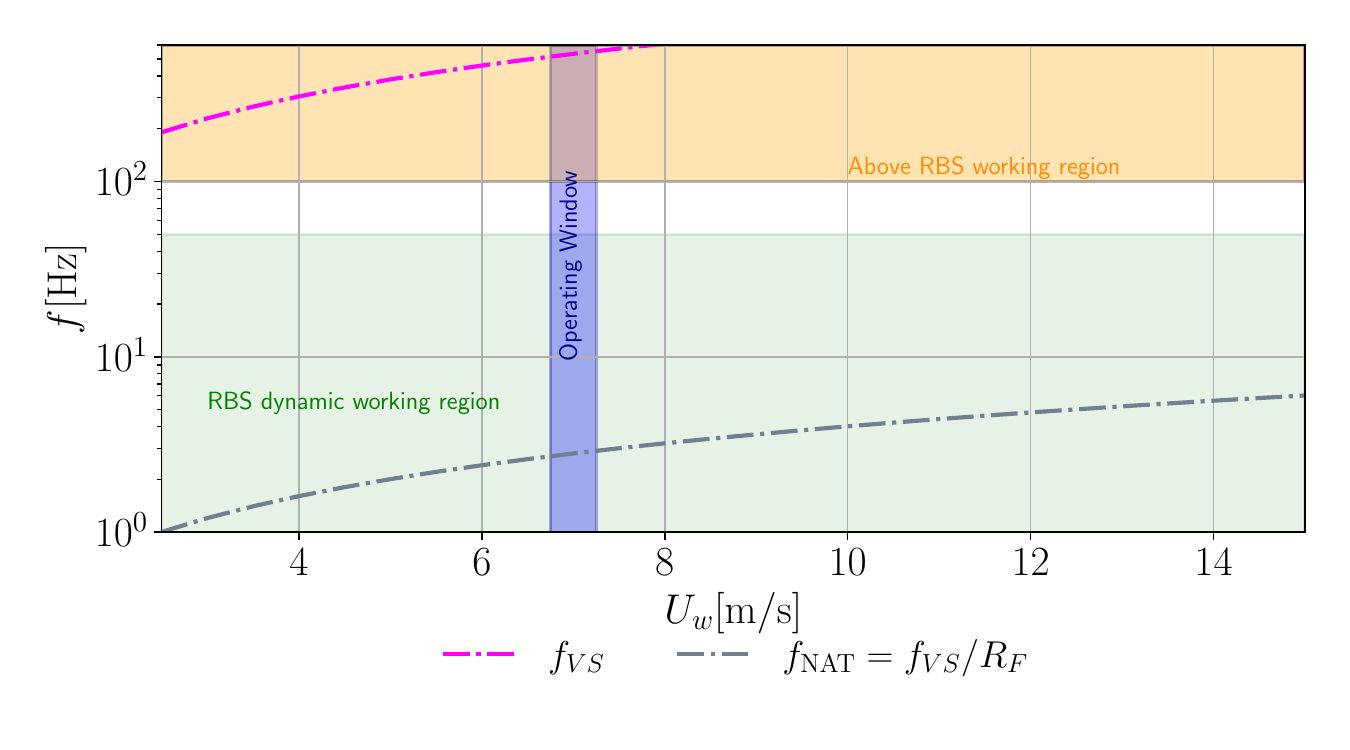}
\caption{Indication of the temporal resolution of the RBS system with respect to the vortex shedding frequency from the conductor $f_{VS}$ and the appropriate natural frequency $f_{NAT}$ as a function of wind speed.}
\label{fig:rbs_fnat}
\end{center}
\end{figure}

Figure \ref{fig:rbs_fnat} illustrates the frequencies of the vortex shedding from the conductor $f_{VS}$ as a function of wind speed.
The natural frequency, determined by setting the tension (sag) in the cable that maintains the correct ratio $R_F$ is also marked onto the figure as well as the temporal resolution of the RBS instrumentation.
It can be seen that at the operating condition $U_w = U(H) \approx 7$ m\,s$^{-1}$, i.e. the mean wind speed incident on the conductor, the RBS is not capable of temporally resolving the vortex shedding (although it is easily capable of resolving the natural frequency of the cable, which is expected to be close to the dominant frequency of aeolian vibration).

\subsection{Test cases}

The wind tunnel was always run such that the wind speed incident on the conductor was 7 m\,s$^{-1}$.
The wind speed was ramped up from zero to the operating condition and data acquisition was only commenced once the pitot-static probe mounted 1 m above the tunnel's floor (that determines the freestream velocity in the wind tunnel) had stabilised for a period of c. 3 minutes.
This was typically c. 5 minutes after tunnel start up.
All data were acquired for a period of 120 seconds which corresponds to c. 360 cycles of the natural frequency of the cable, which is sufficient to ensure converged statistics.

Figure \ref{fig:velprof} illustrates that the wind speed $U$ is a function of height above the ground $y$, i.e. $U = U(y)$.
As the conductor height was varied it was therefore necessary to modify the freestream velocity in the wind tunnel $U_\infty$ to ensure that the mean velocity at the conductor height $U_w = 7$ m\,s$^{-1}$.
Figure \ref{fig:velprof} shows that the turbulence intensity also varies with $y$ so it is important to note that whilst the incident wind speed remains fixed for all test cases the turbulence intensity is largest for the cases where the conductor is closest to the ground ($H = 100$ mm, $H^\star = 0.25$) and smallest when the conductor is at the highest location ($H = 370$ mm, $H^\star = 0.925$).
The various test cases are illustrated in table \ref{tab:testcases}.

\begin{table*}[h!]
\begin{center}
\begin{tabular}{c|c|c|c|c|c|c}
\textbf{Test case} & $H$ (mm) & $H^\star$ & $x_C$ (m) & $U_\infty$ (m s$^{-1}$) & $U_w$ (m s$^{-1})$ & TI (\%) \\ \hline \hline
01 & 100 & 0.25 & - (no turbine) & 11.3 & 7.0 & 21.8 \\ \hline
02 & 100 & 0.25 & 1.5$D$ & 11.3 & 7.0 & 21.8 \\ \hline
03 & 100 & 0.25 & 2$D$ & 11.3 & 7.0 & 21.8 \\ \hline
04 & 100 & 0.25 & 3$D$ & 11.3 & 7.0 & 21.8 \\ \hline
05 & 100 & 0.25 & 4$D$ & 11.3 & 7.0 & 21.8 \\ \hline
06 & 190 & 0.475 & - (no turbine) & 9.39 & 7.0 & 19.1 \\ \hline
07 & 190 & 0.475 & 1.5$D$ & 9.39 & 7.0 & 19.1 \\ \hline
08 & 190 & 0.475 & 2$D$ & 9.39 & 7.0 & 19.1 \\ \hline
09 & 190 & 0.475 & 3$D$ & 9.39 & 7.0 & 19.1 \\ \hline
10 & 190 & 0.475 & 4$D$ & 9.39 & 7.0 & 19.1 \\ \hline
11 & 370 & 0.925 & - (no turbine) & 8.62 & 7.0 & 16.8 \\ \hline
12 & 370 & 0.925 & 1.5$D$ & 8.62 & 7.0 & 16.8 \\ \hline
13 & 370 & 0.925 & 2$D$ & 8.62 & 7.0 & 16.8 \\ \hline
14 & 370 & 0.925 & 3$D$ & 8.62 & 7.0 & 16.8 \\ \hline
15 & 370 & 0.925 & 4$D$ & 8.62 & 7.0 & 16.8 \\ \hline
\end{tabular}
\caption{Various experimental test cases. $H$: conductor height above the ground, $H^\star$: $H/h$ where $h$ is the hub height, $x_C$: turbine -- conductor separation, $U_\infty$: freestream velocity in the wind tunnel, $U_w = U(H)$: incident mean velocity on the conductor, TI: turbulence intensity. Recall $D$ = 0.500 m.}
\label{tab:testcases}
\end{center}
\end{table*}

\section{Results}

\subsection{Time-averaged strains in the cable} \label{sec:meanstrain}

\begin{figure}[h!]
\centering\includegraphics[width=\linewidth]{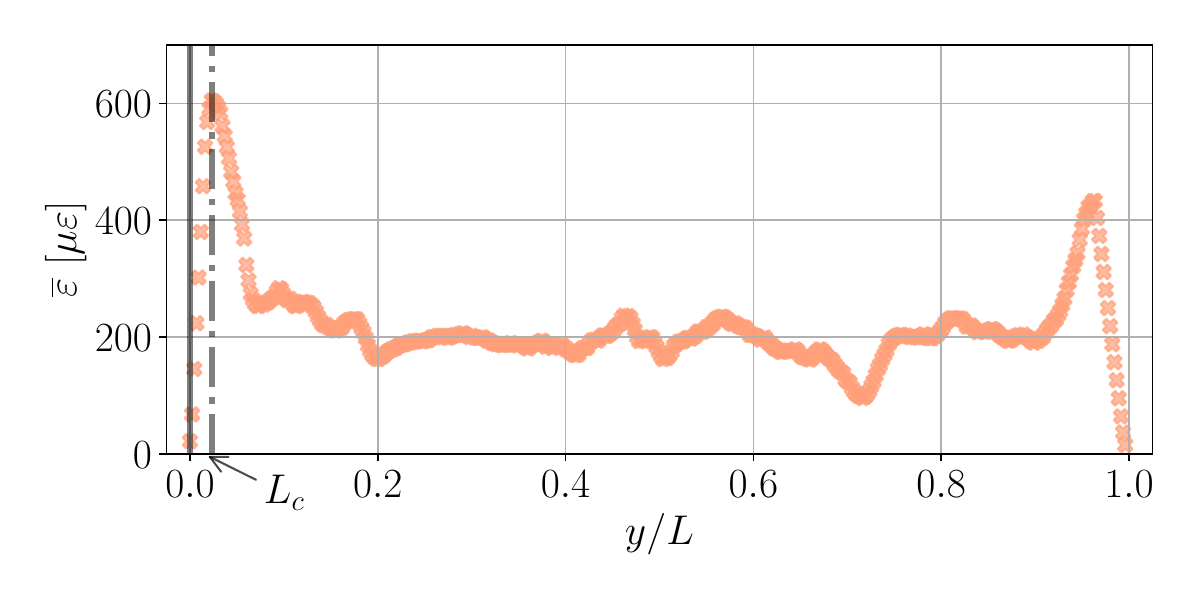}
\caption{The mean strain (in units of micro strain) along the span of the cable in the absence of the wind turbine upstream.
This example, for $H^\star = 0.475$, is illustrative for all three conductor heights.}
\label{fig:meanspan}
\end{figure}

Figure \ref{fig:meanspan} presents the time-averaged (mean) strain within the entire span of the cable.
It is evident that there are two distinct regions within the cable.
The first region is in the central portion of the cable in which the mean strain is approximately constant at $\overline \varepsilon \approx 200 \times 10^{-6}$.
There are small fluctuations about this value which are likely brought about by imperfections in the cable itself, i.e. the cross-sectional area of the cable may slightly vary due to manufacturing imperfections.
The second region, of which there are two instances, is the region immediately adjacent to where the cable is clamped atop the pylon models.
Here the strain adjusts from zero at the location of the clamp to some maximum value, before it falls away to the approximately constant value in the central portion of the cable.
A distance $L_C$ is marked onto the figure.
This is the location at which the mean strain takes on its maximum value and is indicative of the region of the cable that directly feels the influence of the clamp in which it effectively behaves as a vibrating free beam.

\begin{figure*}[h!]
\includegraphics[width=\linewidth]{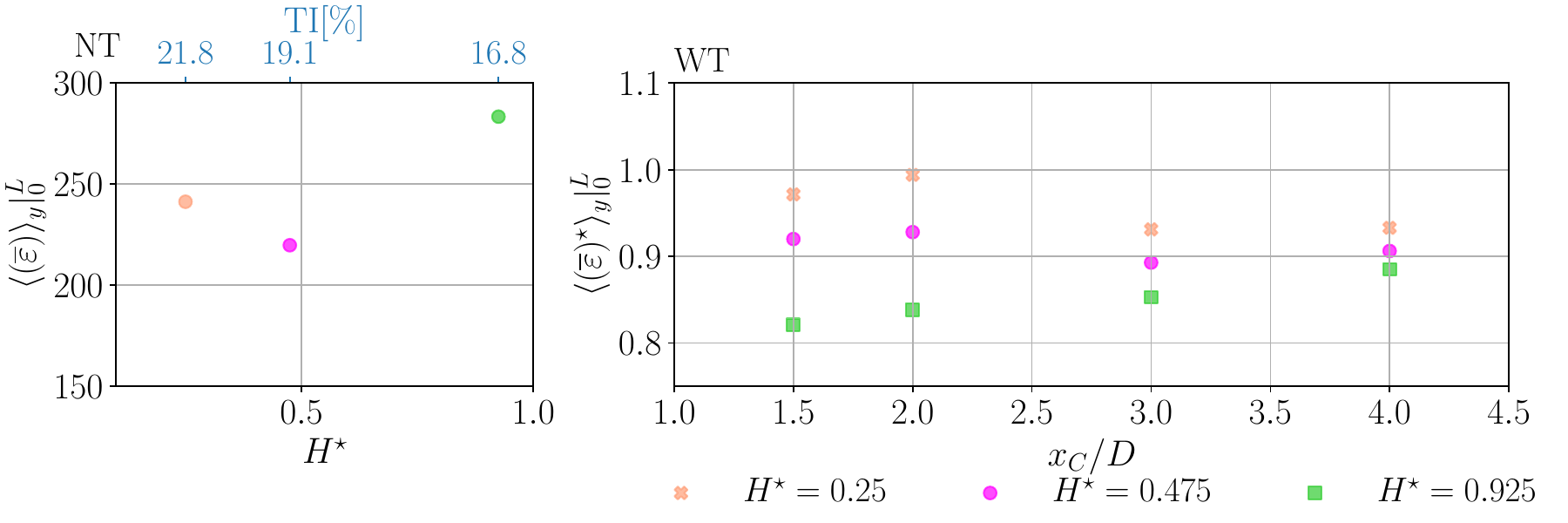}
\caption{The mean strain (in units of micro strain) ensemble averaged across the whole span for the baseline reference where no wind turbine is present (left). 
The mean strain ensemble averaged across the whole span when the turbine model is upstream of the conductor (right). 
The mean strains are normalised with respect to the baseline case so that $\langle (\overline \varepsilon )^\star \rangle _{y} |_0^L < 1$ corresponds to a mean strain that is less than that in the absence of a turbine upstream.}
\label{fig:meanwhole}
\end{figure*}

Two sets of mean strains for the cables are presented in this section.
The first takes the ensemble average of the strain across the \emph{whole span} of the conductor, presented in figure \ref{fig:meanwhole}.
The left hand pane of the figure shows the ensemble averaged strain across the whole span (i.e. averaged in time and along the entire length of the conductor) in the absence of the model wind turbine upstream of the conductor, for the three conductor heights.
It can be seen that the span-averaged strain is highest for the conductor at the greatest height above the ground.
The two lower level conductor heights show similar values of the mean strain, although the intermediate conductor height is lower than the lowest conductor height.

The right hand pane of the figure shows the mean strain ensemble averaged across the whole span of the conductor relative to the no-turbine case, i.e. $\langle (\overline \varepsilon )^\star \rangle _{y} |_0^L < 1$ corresponds to a mean strain that is less than for the reference, no-turbine case.
The headline result is that for all conductor heights, and for all conductor -- turbine separations, the mean strain is always lower with the turbine upstream of the conductor than for the no-turbine case.

There is, however, different behaviour for the three different conductor heights above the ground.
When the conductor is at its highest $H^\star = 0.925$ then it is close to being in the centre of the wind-turbine wake, i.e. close to the hub height (see figure \ref{fig:conheight}).
The velocity deficit $U_d = U_\infty - U$ in a wind-turbine wake decreases with distance downstream from the turbine, i.e. increasing $x$, due to wake recovery brought about by the entrainment and mixing of higher momentum background fluid into the low-momentum wake.
As a result the greater the conductor -- turbine separation then the faster the wind speed becomes that is incident over the conductor due to this wake recovery and hence the greater the induced strain in the cable.
Additionally, the wake spreads out more further downstream (see figure \ref{fig:wake}), again due to entrainment of background fluid.
This means that the wake is incident over a greater proportion of the cable's span which again increases the mean strain ensemble averaged across the whole span.

\begin{figure}[h!]
\includegraphics[width=\linewidth]{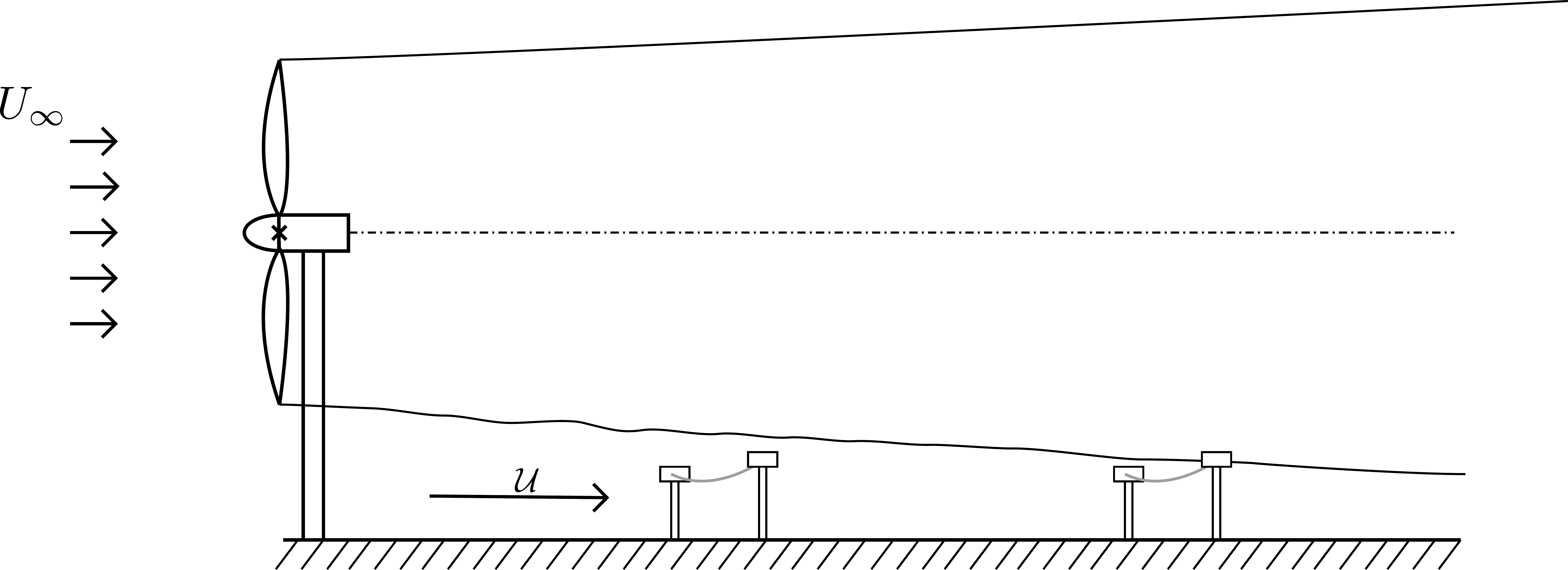}
\caption{Schematic of the proposed mechanism that enhances the mean strain in the conductor for the lowest conductor height $H^\star = 0.25$ when the turbine -- conductor separation is small (i.e. 1.5$D$ or 2$D$).
The presence of the ground and turbine wake ``squeezes'' an accelerated layer of air past the conductor.}
\label{fig:squeeze}
\end{figure}

For the lowest conductor height the conductor is likely situated beneath the turbine wake when the conductor -- turbine separation is $< 3D$.
The presence of this wake, however, is akin to a blockage that will force freestream air to spread out.
The ground acts as an additional blockage and so there is likely a layer of air that is accelerated as it is ``squeezed'' between the turbine wake and the ground which yields larger mean strains for turbine -- conductor separations of $1.5D$ and $2D$.
This postulated mechanism is illustrated in figure \ref{fig:squeeze}.
A similar, but weaker effect is likely happening for the intermediate conductor height $H^\star = 0.475$.
For $x_C/D = 3D$ and $4D$ the wake will have spread sufficiently to be incident on the conductor and we see similar behaviour where the mean strain increases with greater turbine -- conductor separation.
Note that across all $x_c$ locations, including $x_C = 4D$, the mean strain is the lowest for the highest conductor height.
This is likely because the conductor is closer to the centre of the wake, and hence the highest velocity deficit, whereas for the two lower conductor heights the cable is only exposed to the periphery of the turbine wake, with lower associated velocity deficits (higher incident wind speeds).

\begin{figure*}[h!]
\includegraphics[width=\linewidth]{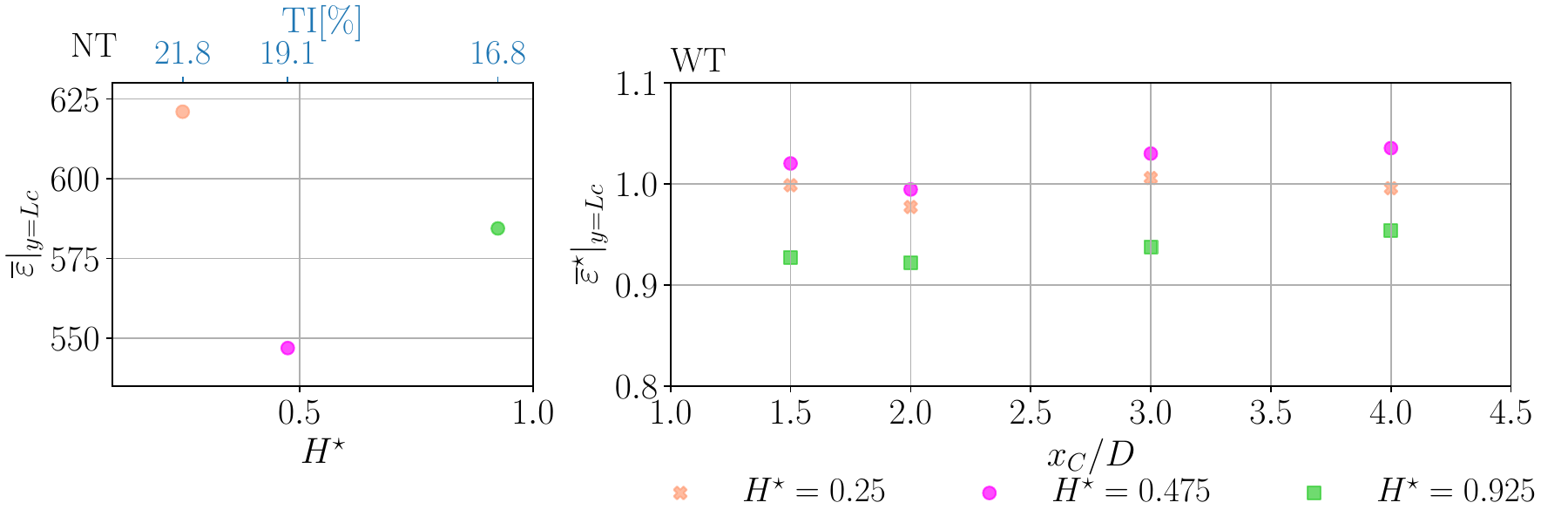}
\caption{The mean strain (in units of micro strain) at the location close to the clamp where the mean strain is maximised, $y = L_C$, for the baseline reference where no wind turbine is present (left). 
The mean strain at $y = L_C$ when the turbine model is upstream of the conductor (right). 
The mean strains are normalised with respect to the baseline case so that $\overline \varepsilon ^\star |_{y=L_C} < 1$ corresponds to a mean strain that is less than that in the absence of the turbine upstream.}
\label{fig:meanpylon}
\end{figure*}

Similar analysis is performed to that presented in figure \ref{fig:meanwhole} at the location along the span where the mean strain is at its maximum ($y = L_C$ in figure \ref{fig:meanspan}).
Due to its exposure to the highest magnitude strains, this is region of the cable that is most likely to experience fatigue failure.
This is presented in figure \ref{fig:meanpylon}.
The left-hand pane is again the baseline case where there is no turbine upstream of the conductor.
It is immediately apparent that the strain magnitudes in this region of the cable are significantly larger than the average values across the whole span.
The highest magnitude strains are this time found in the lowest conductor height $H^\star = 0.25$. 

The right-hand pane again shows the strains for the cases where the turbine was present upstream of the conductor relative to the no-turbine case.
The headline result is again that the presence of a wind turbine upstream of the conductor does not significantly increase the mean strain in this critical region of the conductor, where failure is most likely to occur. 
$\overline \varepsilon ^\star |_{y=L_C}$ remains smaller than unity for the lowest and highest conductor heights and only exceeds unity by around 3\% for the intermediate conductor height.

\subsection{Fluctuating strains in the cables}

\begin{figure*}[h!]
\centering\includegraphics[width=\linewidth]{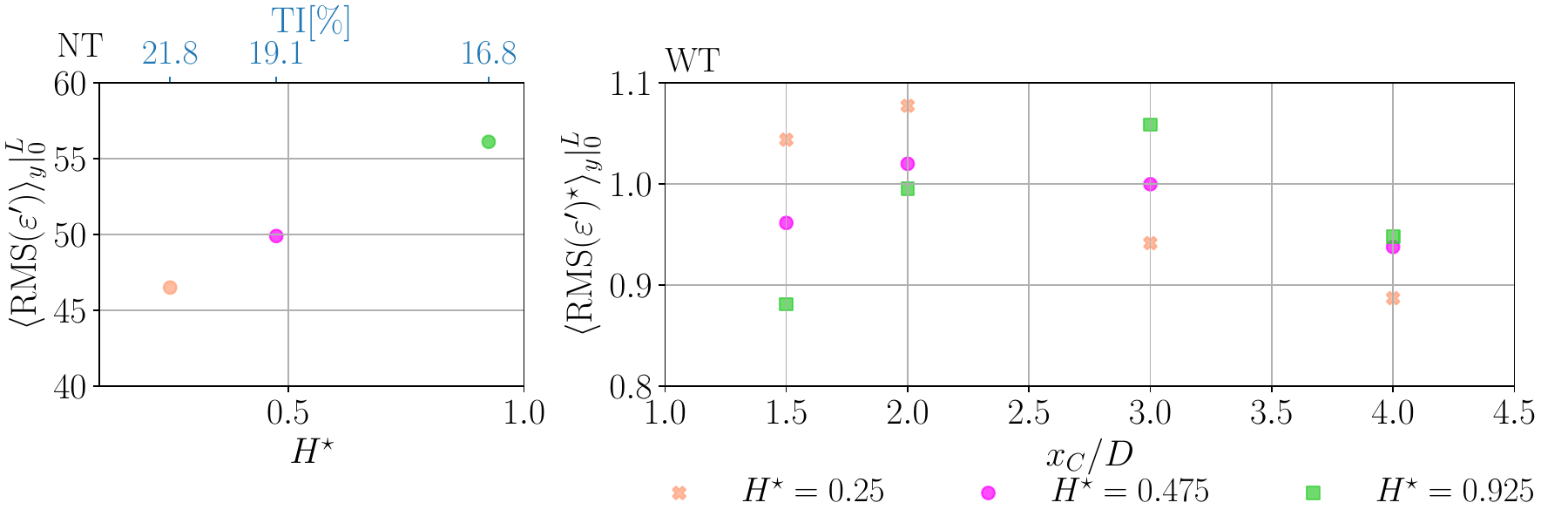}
\caption{The r.m.s. strain (in units of micro strain) ensemble averaged across the whole span for the baseline reference where no wind turbine is present (left). 
The r.m.s. strain ensemble averaged across the whole span when the turbine model is upstream of the conductor (right). 
The r.m.s. strains are normalised with respect to the baseline case so that $\langle \text{RMS}(\varepsilon ' )^\star \rangle _{y} |_0^L > 1$ corresponds to a r.m.s. strain that is greater than that in the absence of the turbine upstream.}
\label{fig:rmswhole}
\end{figure*}

The fluctuating strain $\varepsilon'$ is computed via the Reynolds decomposition
\begin{equation}
\varepsilon' = \varepsilon - \overline \varepsilon
\end{equation}
where $\overline \varepsilon$ is the mean strain, which means that the time average of $\varepsilon'$ is identically zero.
Figure \ref{fig:rmswhole} shows the root mean square (r.m.s.) strain ensemble averaged across the whole span of the conductor with the left-hand pane again being the baseline case with no turbine present and the right pane plotting the RMS of $\varepsilon'$ relative to the baseline case.
We see that the fluctuating strain increases monotonically with conductor height above the ground, but it is noteworthy that the magnitude of the fluctuations is approximately 20\% of the magnitude of the mean strain, ensemble averaged across the whole span (figure \ref{fig:meanwhole}).

When looking at the effect of placing the turbine upstream of the conductor there is a mixed picture.
Only when the turbine is placed 4$D$ upstream of the conductor is the r.m.s. strain fluctuation reduced for all three conductor heights with respect to the baseline.
For the two lowest conductor heights the r.m.s. strain is maximised for $x_C = 2D$, reaching values that are 2\% and 8\% higher than for the baseline case for $H^\star = 0.475$ and $H^\star = 0.25$ respectively.
When the conductor is closer to the centre of the wake this location is pushed further downstream to $x_C = 3D$.
The behaviour of the r.m.s. strain fluctuation for the two closest turbine locations at $H^\star = 0.25$ is more evidence that the ``squeezed'' layer of air between the turbine wake and the ground is augmenting the mechanical response of the conductor.

\begin{figure*}[h!]
\centering\includegraphics[width=\linewidth]{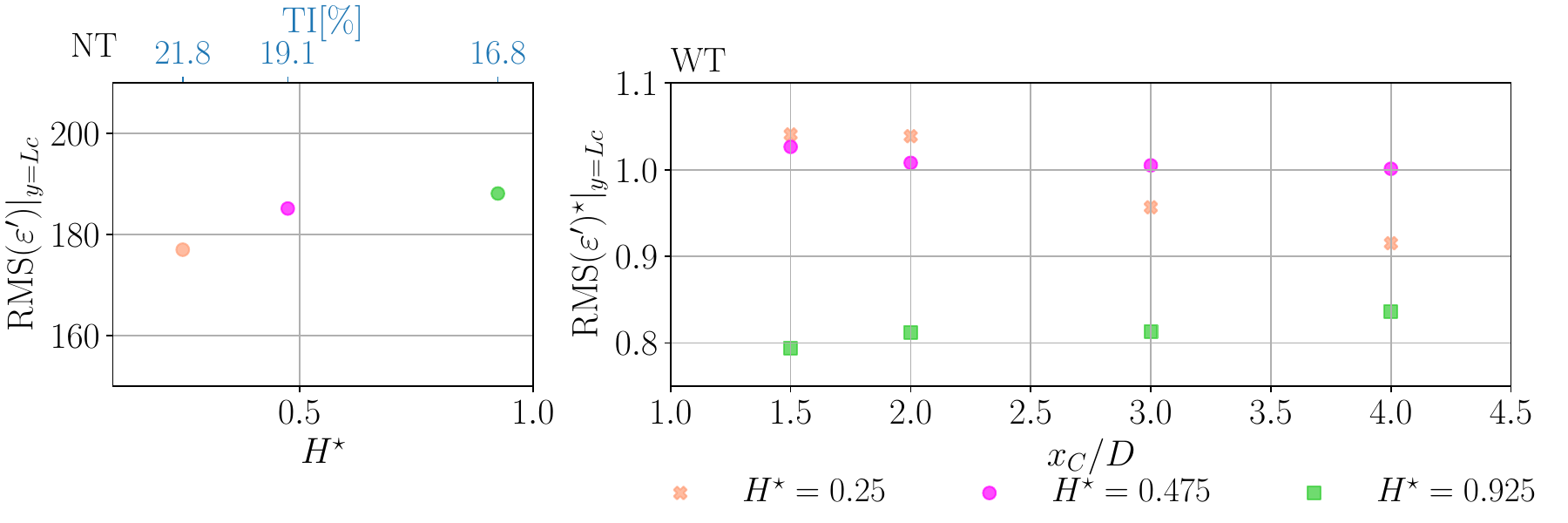}
\caption{The r.m.s. strain (in units of micro strain) at the location close to the clamp where the mean strain is maximised, $y = L_C$, for the baseline reference where no wind turbine is present (left). 
The r.m.s. strain at $y = L_C$ when the turbine model is upstream of the conductor (right). 
The r.m.s. strains are normalised with respect to the baseline case so that $\text{RMS}(\varepsilon ' )^\star |_{y=L_C} > 1$ corresponds to a r.m.s. strain that is greater than that in the absence of the turbine upstream.}
\label{fig:rmspylon}
\end{figure*}

Figure \ref{fig:rmspylon} shows the r.m.s. strain fluctuations at the location where the mean strain is maximised, $y = L_C$.
Comparison of figures \ref{fig:meanpylon} and \ref{fig:rmspylon} shows that for the baseline case (no turbine upstream of the conductor) the r.m.s. fluctuating strain is approximately 28-34\% the value of the mean strain, i.e. both the absolute and relative magnitude of the strain fluctuations are larger at this critical location than they are when ensemble averaged across the entire span.
The baseline r.m.s. strains follow the same trend with conductor height as the r.m.s. strain ensemble averaged across the entire span.

For the intermediate height conductor, $H^\star = 0.475$, the presence of an upstream turbine always marginally increases the fluctuating strain at this critical location, regardless of the turbine -- conductor separation, but only by a maximum of about 3\%.
The lowest conductor height, $H^\star = 0.25$, sees a very slight increase in the fluctuating strain when the turbine is 1.5$D$ or 2$D$ upstream of the conductor (up to around a 4\% increase) but a much more significant reduction when the turbine is 3$D$ or 4$D$ upstream.
For the highest conductor level, $H^\star = 0.925$, the fluctuating strain is always smaller than in the absence of a wind turbine (the baseline), with the fluctuating strain increasing as the turbine moves further upstream.
This is characteristic of being fully immersed in the turbine wake which undergoes wake recovery as the turbine is moved upstream of the conductor.
No such simple explanation exists for the two lower conductor heights in the absence of a more detailed experimental study of the velocity field in the wake/vicinity of the conductor.

\begin{figure*}[h!]
\centering\subfigure[]{\includegraphics[width=0.4\linewidth]{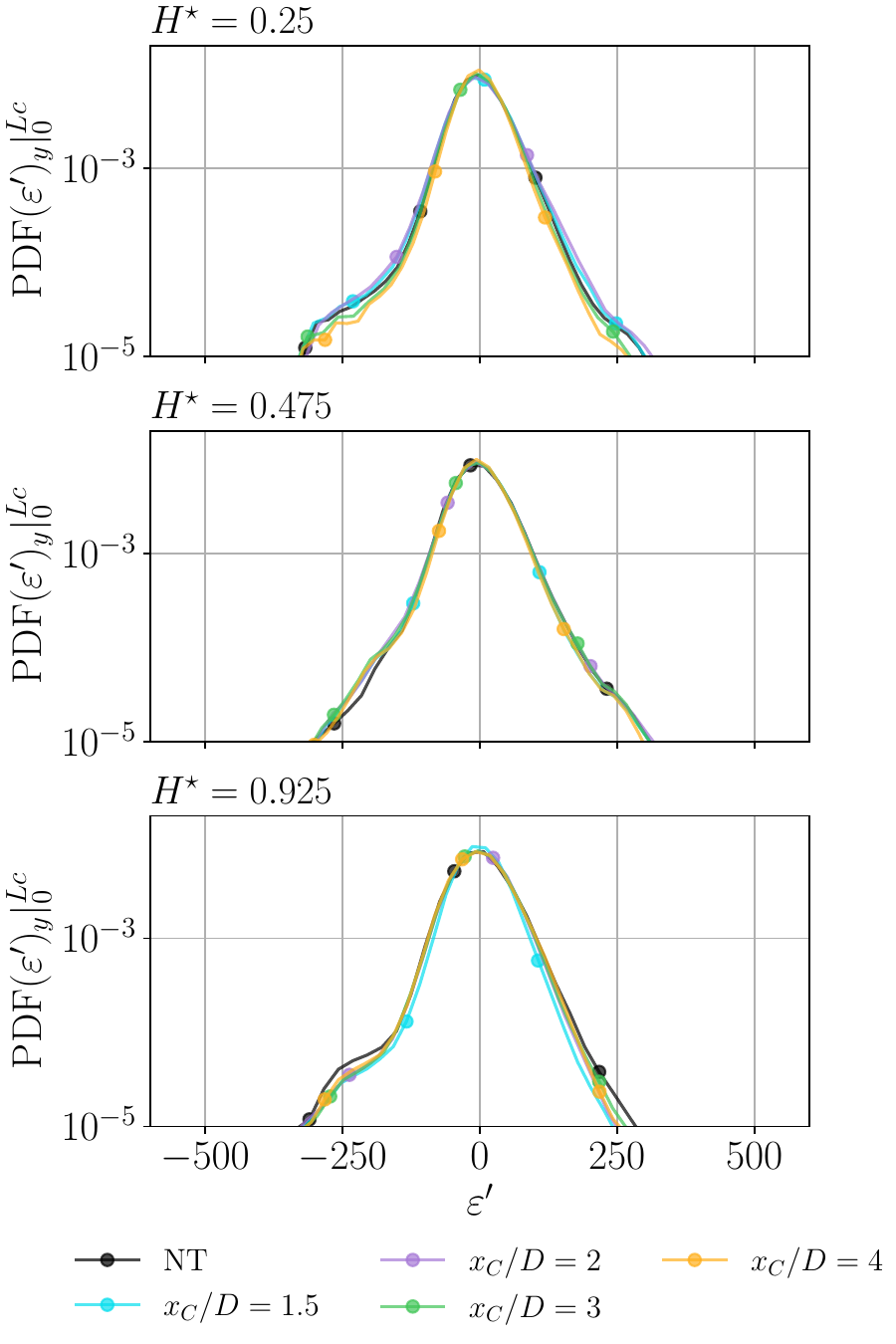}}
\centering\subfigure[]{\includegraphics[width=0.4\linewidth]{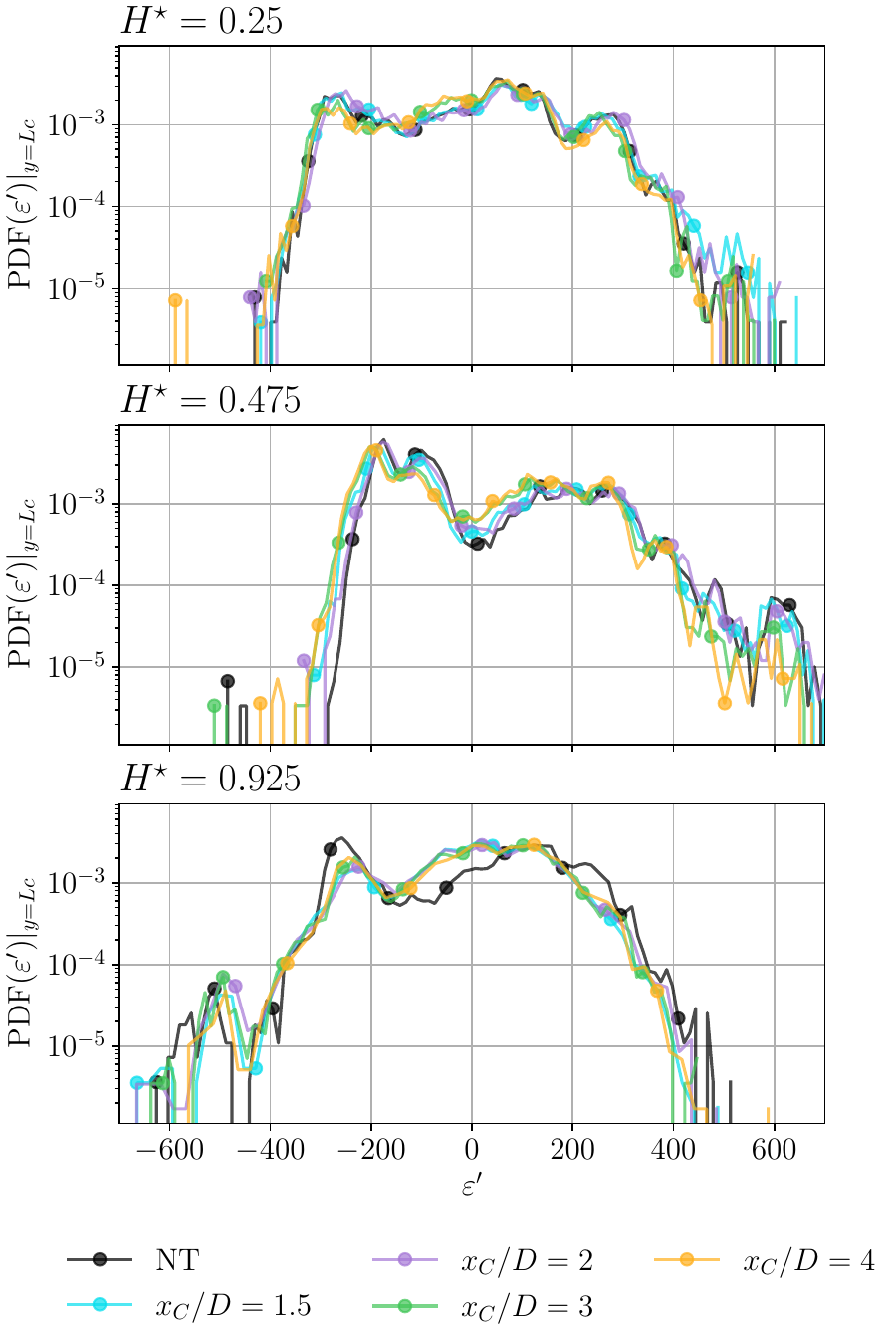}}
\caption{(a) Probability density functions (PDFs) of the fluctuating strain $\varepsilon'$ (in micro strain) computed along the entire span of the cable.
(b) PDF of $\varepsilon'$ (in micro strain) computed at the location of maximum mean strain, $y = L_C$.}
\label{fig:strainpdf}
\end{figure*}

Figure \ref{fig:strainpdf} shows the probability density functions (PDFs) of the fluctuating strain computed along the entire span of the cable (a) and at the location of maximum mean strain, $y = L_C$ (b).
In both cases the PDF for the baseline no-turbine case is plotted in black.
The shapes of the PDFs are very different from one another, with a Gaussian-like distribution for the whole span and a much flatter (and hence more intermittent) distribution at $y = L_C$.
When considering the whole span of the cable it is clear that the presence of an upstream wind turbine has little effect on the distribution of the strain fluctuations.
At $y = L_C$ there is little effect of the upstream turbine at the two lowest conductor heights but if anything the presence of the wind turbine reduces the intermittency of the fluctuating strain (in terms of making the large magnitude events less likely) with respect to the baseline.

\subsection{Spectral analysis of the strain fluctuations}

\begin{figure}[h!]
\centering\includegraphics[width=\linewidth]{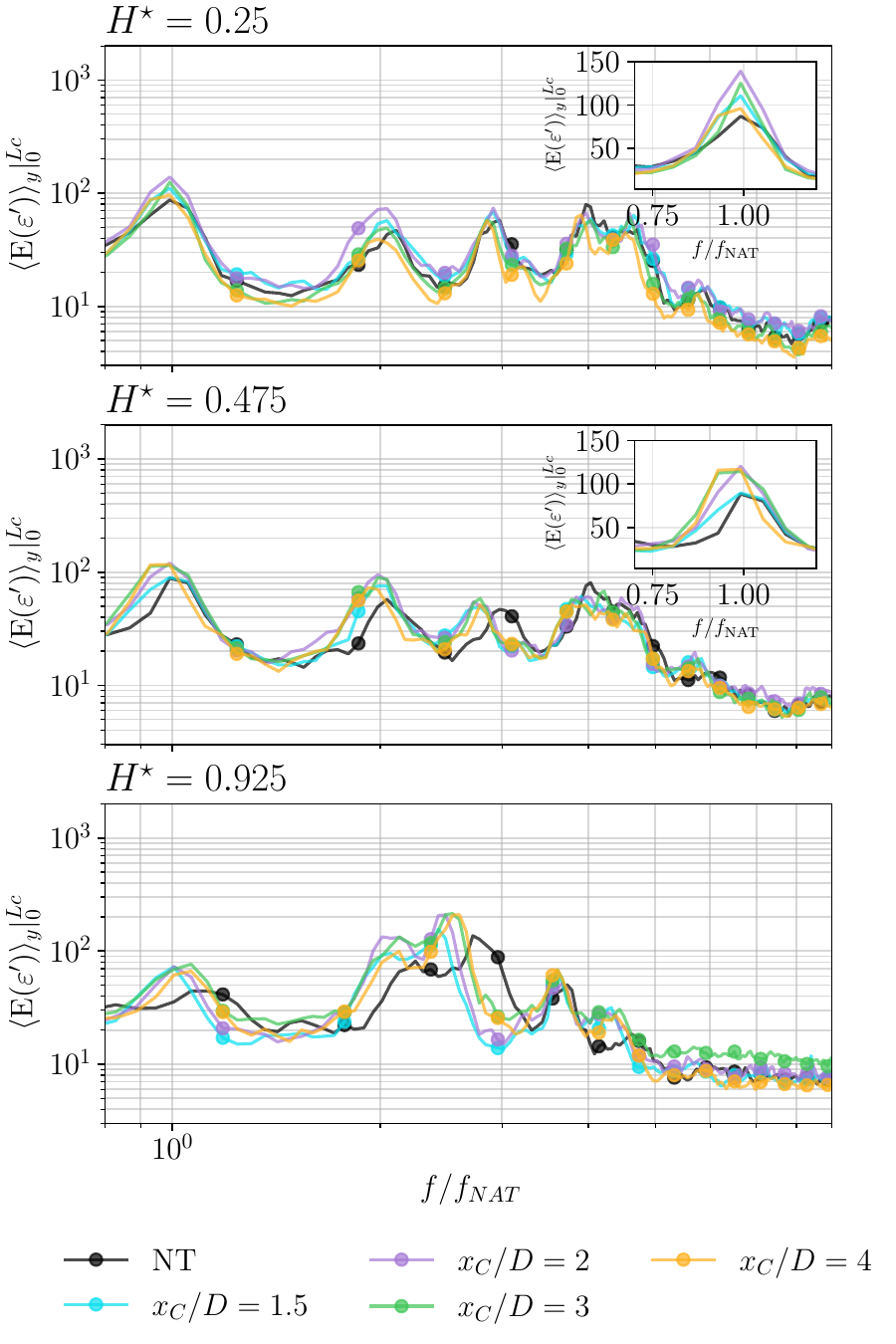}
\caption{Power spectral density of the fluctuating strain within the cable at $H^\star = 0.25$ (top), $H^\star = 0.475$ (middle), and $H^\star = 0.925$ (bottom).
The baseline spectrum (no wind turbine upstream of the conductor) is plotted with the black line.}
\label{fig:mean_spectrum}
\end{figure}

The spectral content of $\varepsilon '$, ensemble averaged across the whole span of the conductor, is presented in figure \ref{fig:mean_spectrum}.
The frequency axis is normalised by the natural frequency of the cable $f_{NAT}$.
We see that for the two lowest conductor heights the largest spectral peak occurs at the natural frequency.
This is indicative of aeolian vibration whereby the aerodynamics (vortex shedding) lead to an excitation of the conductor at its natural frequency.
We also observe harmonics of the natural frequency, i.e. $2f_{NAT}$ and $3f_{NAT}$.
At the lowest conductor height $H^\star = 0.25$ we see that the location of the spectral peaks remains at the same frequency regardless of the turbine -- conductor separation.
This is likely because the conductor is too low to be directly exposed to the wake of the turbine so it experiences a wind speed that is similar to the no-turbine case.
The presence of an upstream wind turbine increases the amplitude of the spectral peak at $f = f_{NAT}$, regardless of how far upstream the turbine is (within the range tested).
A turbine -- conductor spacing of 2$D$ has the biggest effect on the aeolian vibration with the peak being approximately double that for the no-turbine case (see the inset of the top pane of the figure).
Nevertheless, the presence of an upstream turbine increases the amplitude of the spectrum for $f = f_{NAT}$ regardless of the turbine -- conductor spacing.
The picture is more complicated for the harmonics, with seemingly a turbine -- conductor spacing of 2$D$ having the largest effect (increasing the amplitude) and a reduction in the amplitude of the spectral peak for $x_C = 4D$ relative to the no-turbine baseline.

Contrastingly, when the turbine is upstream of the conductor at its intermediate height $H^\star = 0.475$ the spectral peaks shift to lower frequencies (both for the fundamental mode and the harmonics) relative to the no-turbine baseline.
This is indicative of the fact that when exposed to the slower wind within the turbine wake the vortex shedding frequency will decrease (in order to preserve a similar Strouhal number).
It is also evident that the magnitude of these peaks increases relative to the baseline no-turbine case when the wind turbine is present upstream of the cables yielding evidence that the turbine enhances the amplitude of aeolian vibration.

At both of the two lowest conductor heights there is little difference between the spectra at higher frequencies, away from the fundamental mode and harmonics of the natural frequency.
This portion of the spectrum is likely driven by turbulent buffeting of the cable, as opposed to aeolian vibration.
It thus seems that the presence of a wind turbine upstream of the conductor has little effect on the turbulent buffeting, which in any case is over one order of magnitude smaller than the strain fluctuations contained within the natural frequency band of the cable.

The bottom pane of figure \ref{fig:mean_spectrum} shows the spectra for the highest conductor location, $H^\star = 0.925$.
This picture is harder to interpret since the major spectral peak is no longer at $f_{NAT}$.
Instead there is a broadband ``hump'' in the spectrum between the first and second harmonics of $f_{NAT}$, which shifts to lower frequencies when the wind turbine is placed upstream of the conductor.
This is further evidence that the slower wind speed within the turbine wake leads to a lower frequency excitation of the conductor.
The presence of the wind turbine upstream of the conductor also increases the amplitude of the spectral peaks, as occurred with the two lower conductor heights.

One possible explanation for this lies in the fact that a spike at $f_{NAT}$ is evident when the wind turbine is placed upstream of the conductor but it is not present in the baseline no-turbine case.
At this conductor height the turbulence intensity is the lowest so it is possible that these particular turbulent conditions lead to incoherent aerodynamic vortex shedding which does not excite a coherent response from the conductor.
When the turbine is placed upstream of the conductor then there is additive turbulence intensity in the wake.
This brings the turbulence intensity incident upon the conductor to be in-line with the baseline no-turbine cases for the lower conductor heights, which are both shown to elicit a coherent response from the conductor at $f_{NAT}$.
Our previous work has shown that enhanced turbulence intensity can increase the spanwise correlation of vortex shedding from a cylinder thereby eliciting an enhanced mechanical response \citep{oliveira25}.

Figure \ref{fig:rmswhole} shows that for $H^\star = 0.925$ the r.m.s. of the strain fluctuation is maximised for a turbine -- conductor spacing of 3$D$.
The spectrum shows that there is little difference in the magnitudes in the peaks at the natural frequency and in the broadband ``hump'' with respect to other turbine -- conductor spacings, but in fact the major difference appears to be at the high frequencies where the green line clearly sits above the others.
This suggests that there is a strong contribution to turbulent buffeting for this particular configuration.
A location of 3$D$ is typically considered to be the location at which the transition from near-wake to far-wake dynamics occurs (along the wake centre-line) which may explain the spike in turbulent buffeting for this case.
However, more evidence in the form of velocity-field information in the vicinity of the conductor is required in order to test this speculation.

\subsection{Fatigue analysis}

The Poffenberger-Swart (PS) formula provides an estimate of the bending stress range at a cable support due to vibration \citep{poffenberger1965}. 
It assumes that the cable vibrates in a mode shape that creates a relative motion between the conductor and the support point. 
To estimate the number of cycles at a given bending stress a conductor can endure before failure occurs, the idealised bending stress of the conductor from the bending amplitude at its surface, can be measured by utilising the Poffenberger-Swart formula
\begin{equation}
    \sigma_{a}=KY_{b}. \label{eq:sigm_a}
\end{equation}
$Y_b$ is the peak-to-peak bending amplitude and $K$ is the Poffenberger-Swart constant, which is used to convert the measured bending amplitude to the idealised bending stress 
\begin{equation}
    K=\frac{\frac{dp^2E}{4}}{e^{-py}-1+py}.
\end{equation}
Here $d$ is the diameter of the conductor in mm, $E$ is the Young's modulus of the cable in N\,mm$^{-2}$, $y$ is the distance of the point of measurement of strain from the last point of contact between the clamp and conductor in mm ($y = L_C$ is of particular interest), and $p$ is found from the following expression
\begin{equation}
    p=\sqrt{\frac{T}{EI}} .
\end{equation}
Here, $T$ is the conductor tension in N and $EI$ is the flexural rigidity of the cable in N\,mm$^{-2}$. 
For our experiment, $EI$ is calculated from
\begin{equation}
    EI=E\frac{\pi d^4}{64}
\end{equation}
with $E$ being the Young's modulus of the cable in N\,mm$^{-2}$. 
In the absence of the material parameter $E$ from the conductor provider, i.e. EPDM cable, this value was estimated following the semi-empirical relationship between Shore hardness and elastic modulus \citep{gent1958,meththananda2009} to be $E=$ 5.54 N\,mm$^{-2}$. 

The bending amplitude $Y_{b}$ is defined as the peak-to-peak vertical displacement of the conductor. 
In our experiment the strain is measured with an RBS sensor, with a spatial resolution of 2.6 mm. 
Therefore, the bending amplitude $Y_{b}$ can be directly approximated using the distributed strain output over length $L_C$. 
Following this, the bending stress at the critical location ($L_{C}$) can be approximated using \eqref{eq:sigm_a}.

This bending stress history contains fluctuations representing aeolian excitation and turbulent buffeting, and is therefore processed using rainflow counting \citep{amzallag1994}, which decomposes the non-periodic stress signal into a set of individual stress-range cycles. 
These cycles are typically parameterised in terms of the stress jump ($\Delta \sigma = \sigma_{max} - \sigma_{min}$) or mean stress ($\overline \sigma = \frac 12 [\sigma_{max} + \sigma_{min}]$) over which the cycle takes place.
Each extracted cycle represents one bending event contributing to fatigue damage. 
The resulting rainflow histogram provides the number of occurrences of each stress-range magnitude, which can then be combined with an appropriate stress --  life-cycle (S-N, or W\"{o}hler) curve for the conductor material to evaluate the allowable number of cycles at each range before fatigue failure is deemed to have occurred. 
Generally, by applying Miner's cumulative damage rule, the ratio of counted cycles to allowable cycles is summed across all stress ranges to yield the total fatigue damage index. 
In this way, rainflow counting transforms the measured dynamic stress response of the overhung cable into a quantitative estimate of its fatigue life under the observed vibration conditions. 
Given that an EPDM cable is not truly representative for the stress levels of an in-service overhead conductor (OHC), cumulative absolute damage calculation is meaningless.
We thus instead present relative damage metrics, $D_{rel}$, to determine which scenario is more damaging, in which the turbine-present test cases are compared to the baseline no-turbine case.

In fatigue analysis, the higher stress range cycles are more damaging motivating their representation in the damage metric through a power law, e.g. Paris law in which the constants are material parameters that are determined experimentally.
Accordingly, the proposed cumulative damage metric is presented in the form of a power law, where the exponent $m$ effectively weights the influence of higher stress cycles to the overall fatigue life of the conductor
\begin{equation}
    D_{rel}= \sum_{i=1} n_i (\Delta \sigma_b)^m  \label{eq:DREL} .
\end{equation}
Here $m$ is an exponent that varies depending on the material undergoing fatigue cycles, which for our test case is unknown and is therefore approximated. 
The S-N data sets in the literature \citep[e.g.][]{ruellan2019} clearly show much flatter S-N curves for rubber than metals, and linear log-log fits for specific compounds often give exponents around 1-2.
We therefore choose to conduct the analysis for two values of $m$, 1.5 and 2.
$m=$ 1.5 lies in the middle of the suggested range for elastic materials whilst 2 is right at the upper extreme.
Consideration of \eqref{eq:DREL} shows that increasing $m$ has the effect of more heavily weighting the contribution to the cumulative damage from high-magnitude bending stress events and so we can thus think of the $m =$ 2 case as being indicative of the sensitivity of the cable to fatigue failure due to high-magnitude bending stress.
In both cases a quantitative comparison of the incurred damage metric relative to the baseline is presented in this section.
Note, however, that for metals (e.g. in-service conductors) much higher exponents are typically used e.g. $m =$ 3.59 in standard aerospace applications \citep{zhuang2024}.

\subsubsection{Bending stress calculation}
To calculate the bending stress $\sigma_b$, the bending amplitude $Y_b$ is first approximated from the RBS sensor output. From the clamping region up to $L_C$, the OHC behaves as a free-end vibrating beam. 
The cable’s shape can thus be reconstructed from the measured strain using classical beam theory. 
The incremental rotation $\beta_n$, and deflection $\delta_n$ at each sensorised segment are obtained recursively as
\begin{align}	
	\begin{split}
		\beta_n &= \frac{2\Delta L \times \varepsilon_{n-1}}{l_t} + \sum_{j=1}^{n-1}\beta_j , \\
		\delta_{n} &= \Delta L \times \left(\frac{2\Delta L\times\varepsilon_{n-1}}{l_t} + \beta_{n-1}\right) + \delta_{n-1},  
	\end{split}
	\label{eq:shape_sensing_algorithm}
\end{align}
where $\beta_n$ and $\delta_n$ denote, respectively, the rotation and transverse deflection at node $n$. 
The global deflection profile is obtained by integrating the locally measured strain increments together with the corresponding rotations, as defined in \eqref{eq:shape_sensing_algorithm} \citep{oliveira25}. 

\begin{figure}[h]
\centering\includegraphics[width=\linewidth]{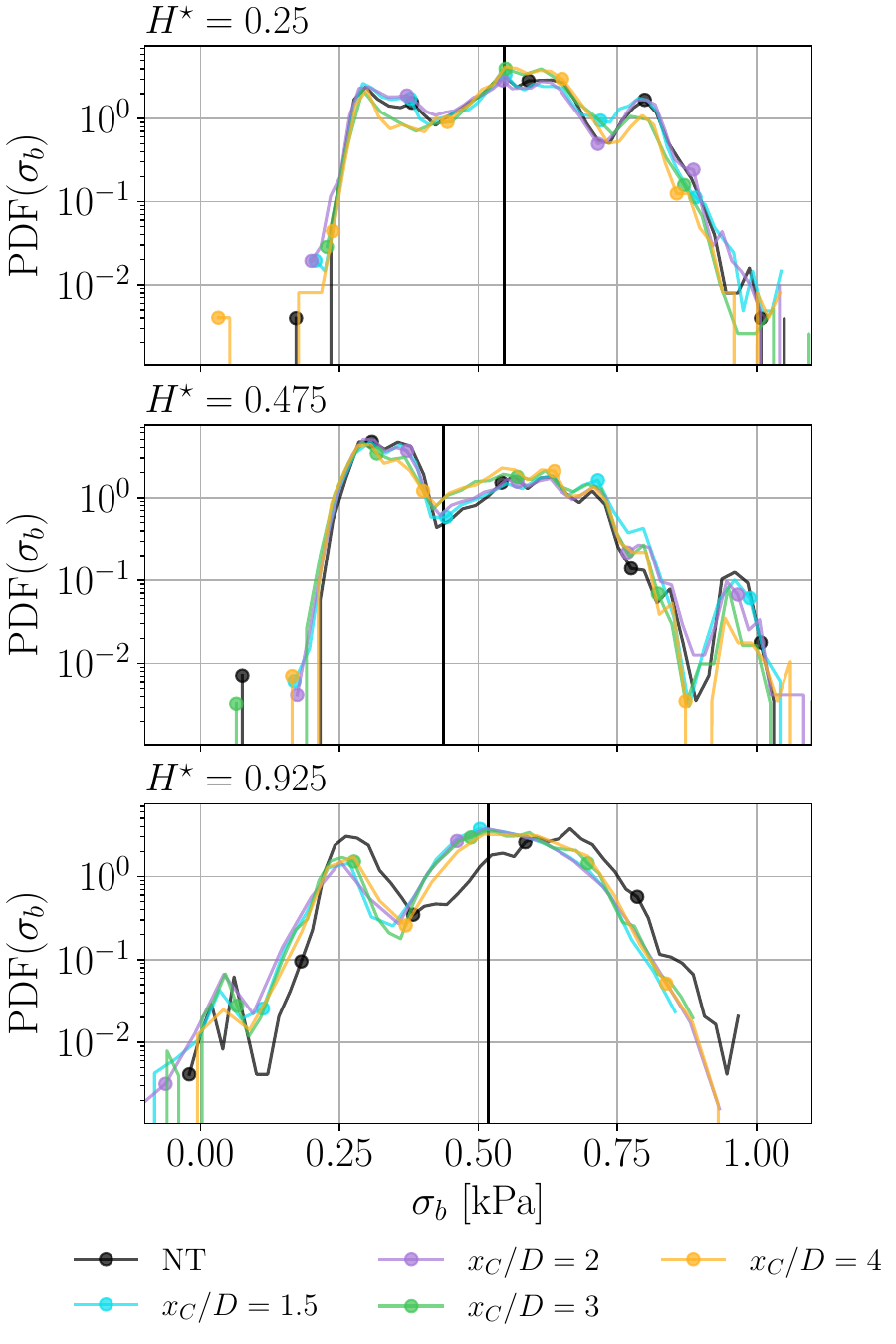}
\caption{Probability density functions (PDFs) of the bending stress $\sigma_b$ of the cable at the location with maximum mean strain $y = L_C$.
The mean bending stress is marked with a solid black line for reference.}
\label{fig:defpdf}
\end{figure}
The bending amplitude $Y_{b}$ is defined as the peak-to-peak vertical displacement of the conductor. Accordingly, $Y_{b}=\delta(y=L_{C})$ in our experiment. 
Following \eqref{eq:sigm_a} the bending stress $\sigma_{b}$ is calculated for each test case.  
\subsubsection{Results and discussion}
Different plots are presented in this section to analyse the effect of the turbine location and OHC height on the accumulation of fatigue damage in the conductor, starting with the bending stress variability at $L_C$. 
Figure \ref{fig:defpdf} shows the probability density functions for the bending stress $\sigma_b$ at location $y = L_C$, the location where the mean strain is maximised.
The PDFs are multi-modal and also significantly flatter than a Gaussian which is indicative of intermittent behaviour.
It is possible that the various modes in the PDFs correspond to the fundamental modes/harmonics of the natural frequency of the cable, i.e. are directly associated to aeolian vibration.
The mean stress is marked onto the figures with a solid black line for reference.

\begin{figure*}[h!]
\subfigure[]{\includegraphics[width=\linewidth]{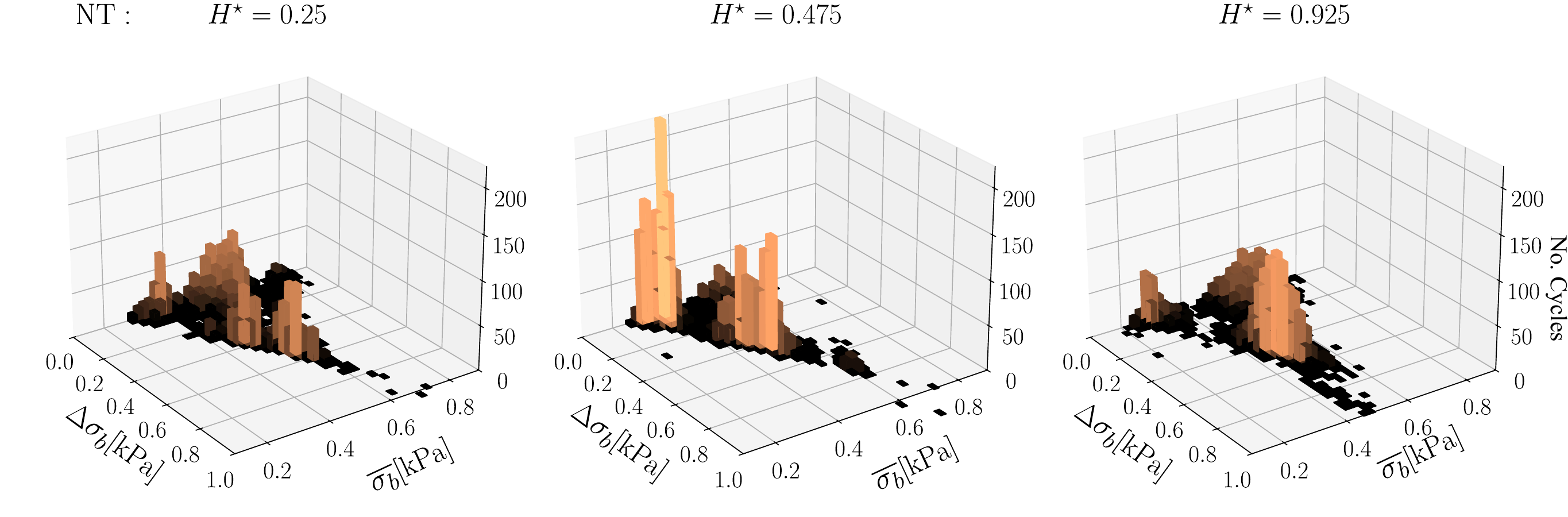}}
\subfigure[]{\includegraphics[width=\linewidth]{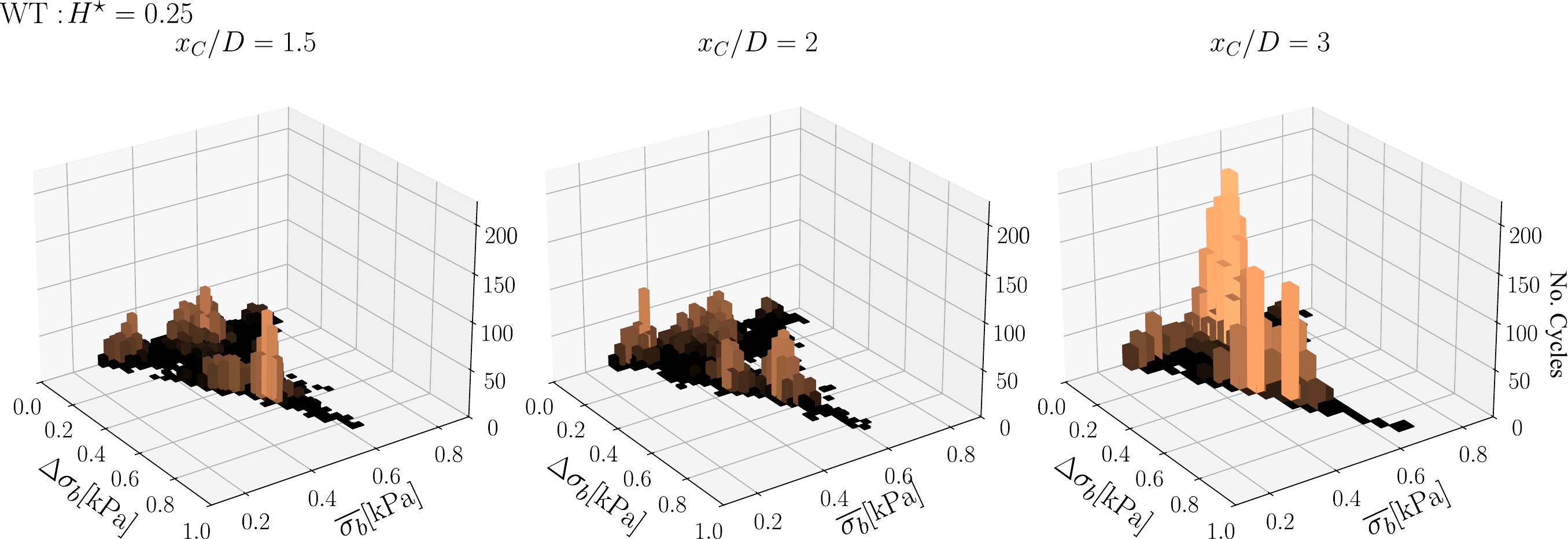}}
\caption{Rainflow counting in the two-dimensional parameter space $\{\Delta \sigma_b, \overline \sigma_b\}$ at $y = L_C$ for (a) the baseline no-turbine cases and (b) various cases with a turbine upstream of the conductor at the lowest height above the ground $H^\star = 0.25$.}
\label{fig:3drainflow}
\end{figure*}

\begin{figure*}[h!]
\begin{center}
\subfigure[]{\includegraphics[width=0.32\linewidth]{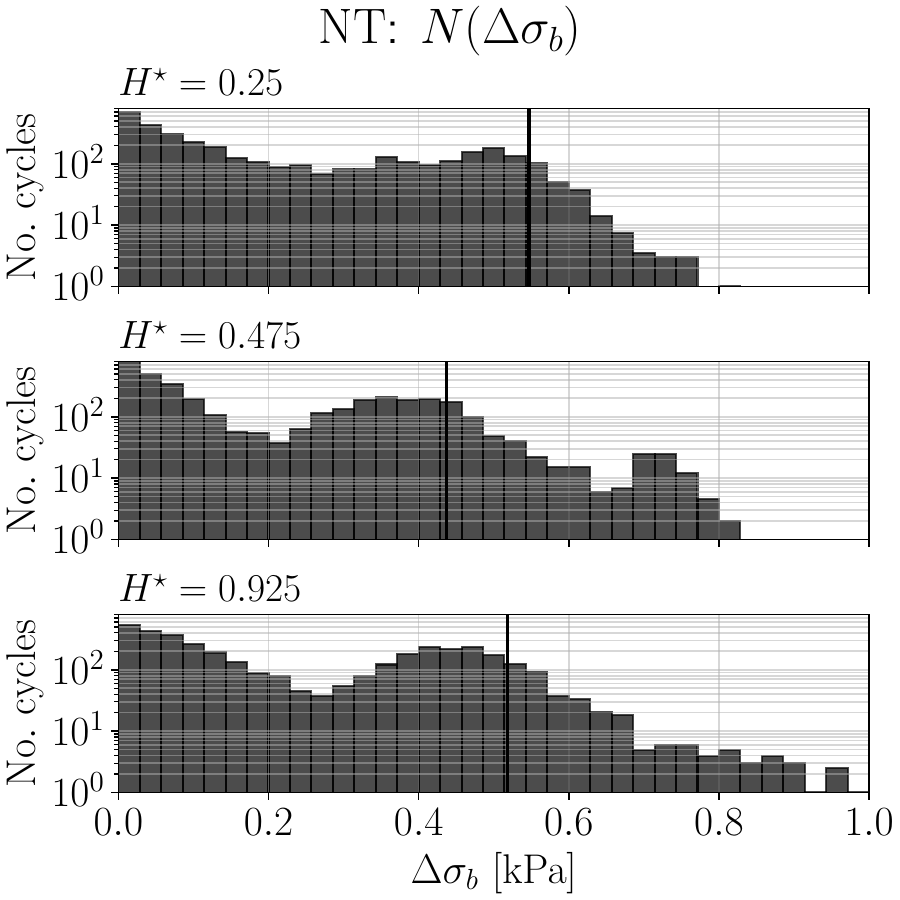}}
\subfigure[]{\includegraphics[width=0.32\linewidth]{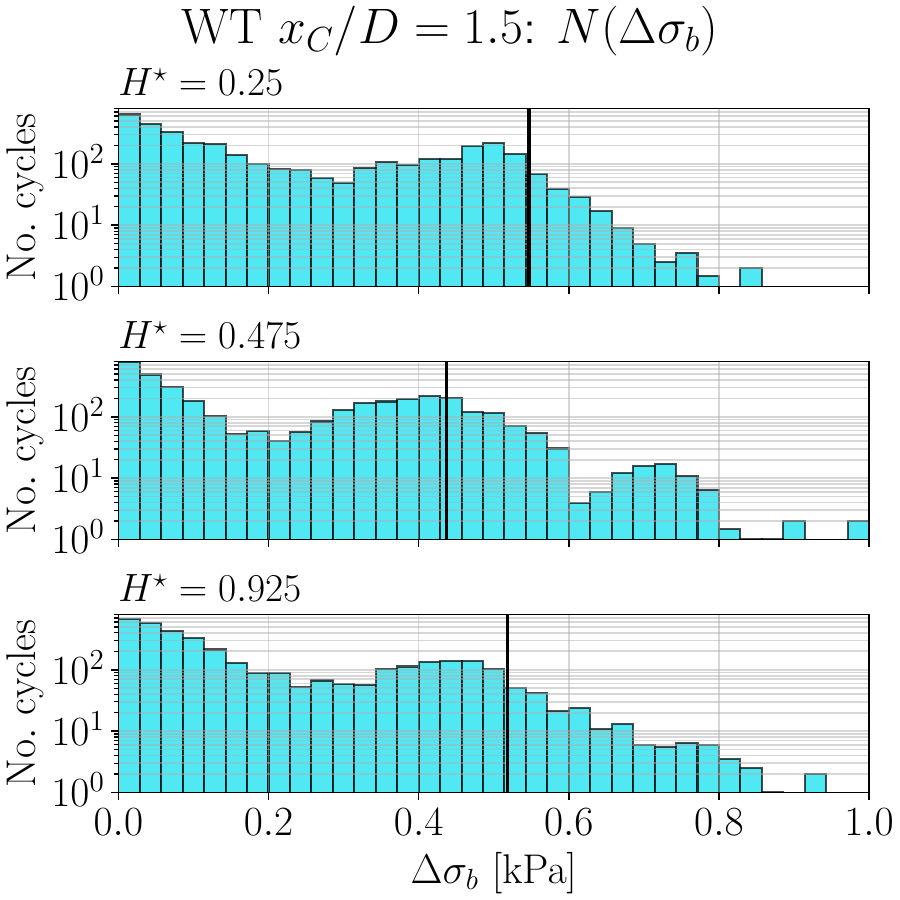}}
\subfigure[]{\includegraphics[width=0.32\linewidth]{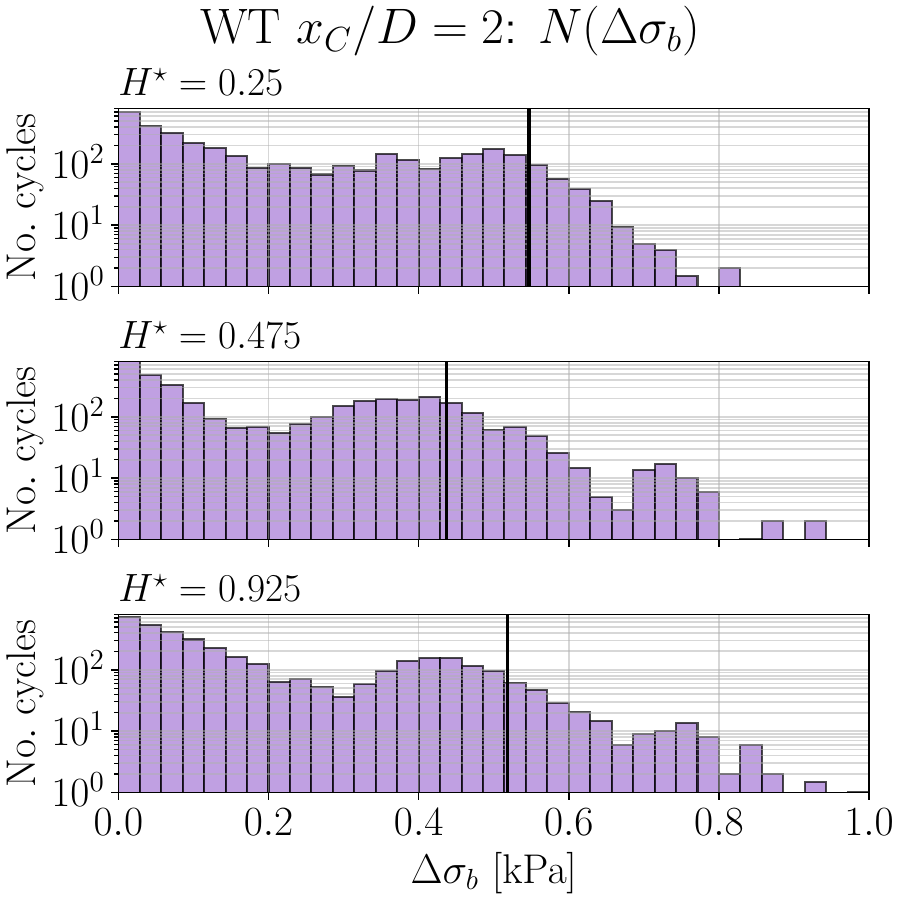}}
\subfigure[]{\includegraphics[width=0.32\linewidth]{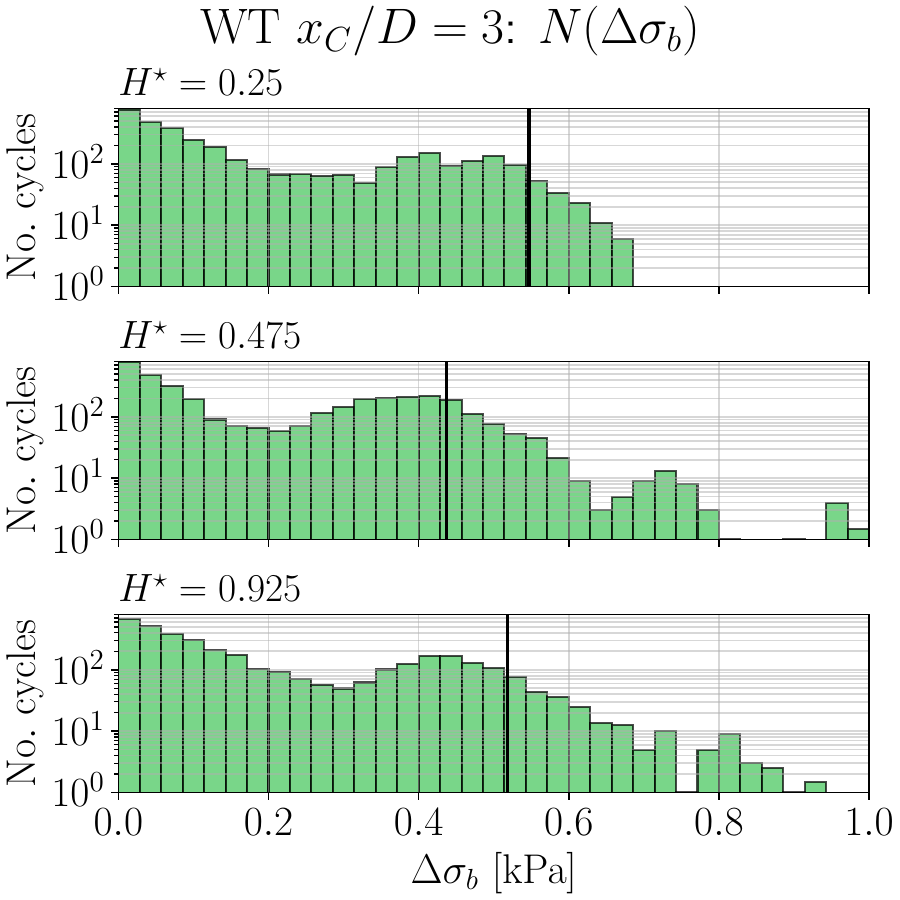}}
\subfigure[]{\includegraphics[width=0.32\linewidth]{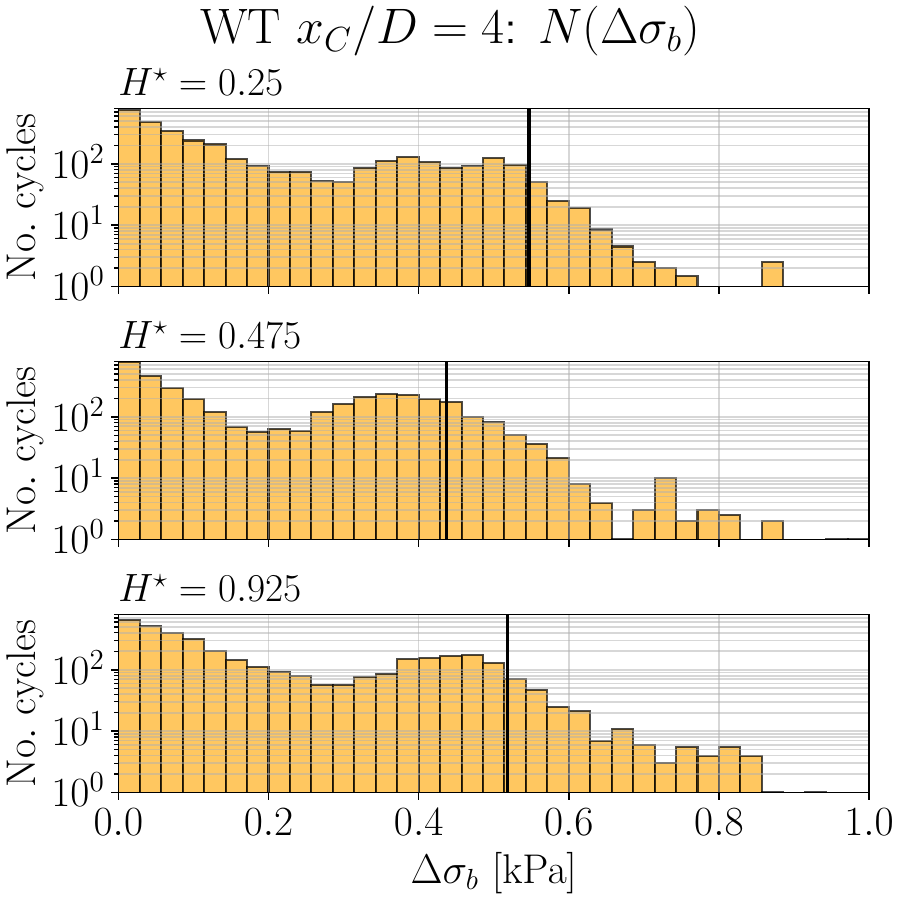}}
\end{center}
\caption{Rainflow counting in the one-dimensional parameter space $\Delta \sigma_b$ of the cable at $y = L_C$. 
(a) the baseline no-turbine cases, (b) turbine -- conductor spacing $x_C =$ 1.5$D$, (c) $x_C =$ 2$D$, (d) $x_C =$ 3$D$, and (e) $x_C =$ 4$D$.} 
\label{fig:2drainflow}
\end{figure*}

Rainflow plots of all test scenarios are presented to demonstrate the change in occurrence of all stress cycles to which the conductor is exposed, and the likely effect on the fatigue lifetime, when comparing the presence of the wind turbine upstream of the conductor to the no-turbine case.
Figure \ref{fig:3drainflow} presents the rainflow counting analysis in the two dimensional parameter space $\{\Delta \sigma_b, \overline \sigma_b\}$ at $y = L_C$  for the baseline no-turbine case (a) and various turbine -- conductor separation cases for $H^\star = 0.25$ (b).
Whilst consideration of the two-dimensional parameter space is more rigorous it is difficult to interpret the plots motivating figure \ref{fig:2drainflow} which presents the rainflow counting analysis only with respect to $\Delta \sigma_b$.
The mean stress is again marked with solid black lines for reference.
The figures all show a second maximum for stress between 0.3 and 0.45 kPa.
This is suggestive of a stress event that occurs regularly and within a particular range and is likely caused by aeolian vibration of the cable since we would expect turbulent buffeting to be more broadband in nature.

\begin{figure*}[h!]
    \centering
    \raisebox{1.5in}{a)}\includegraphics[width=0.9\linewidth]{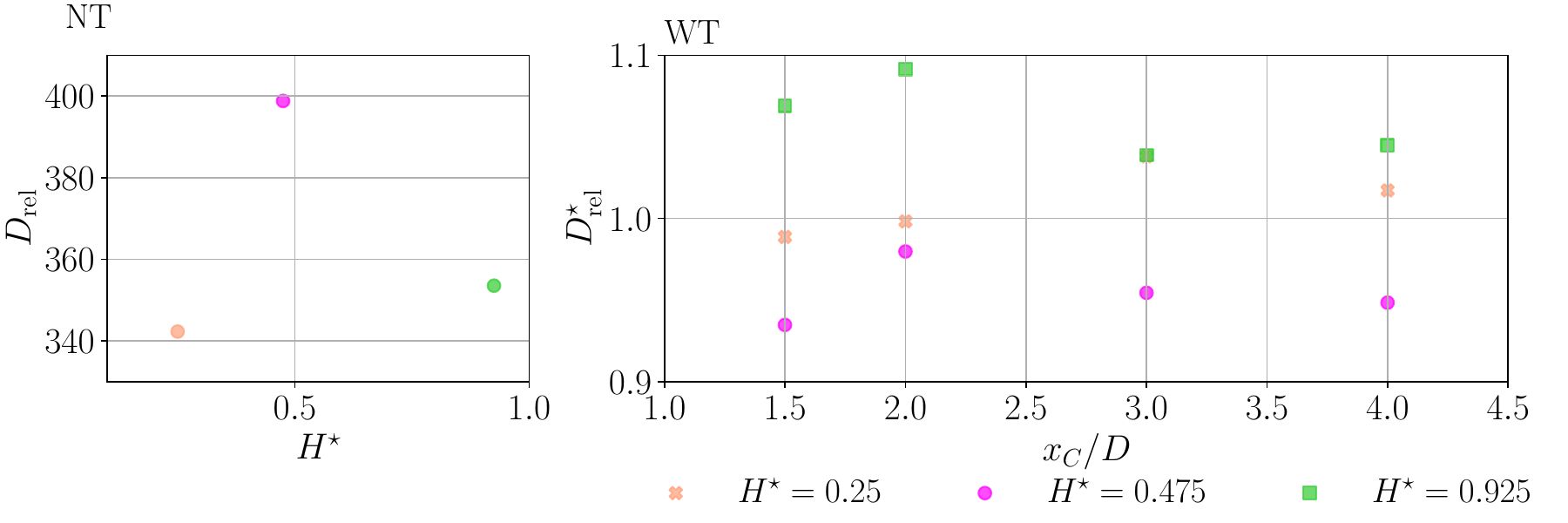}
    \raisebox{1.5in}{b)}\includegraphics[width=0.9\linewidth]{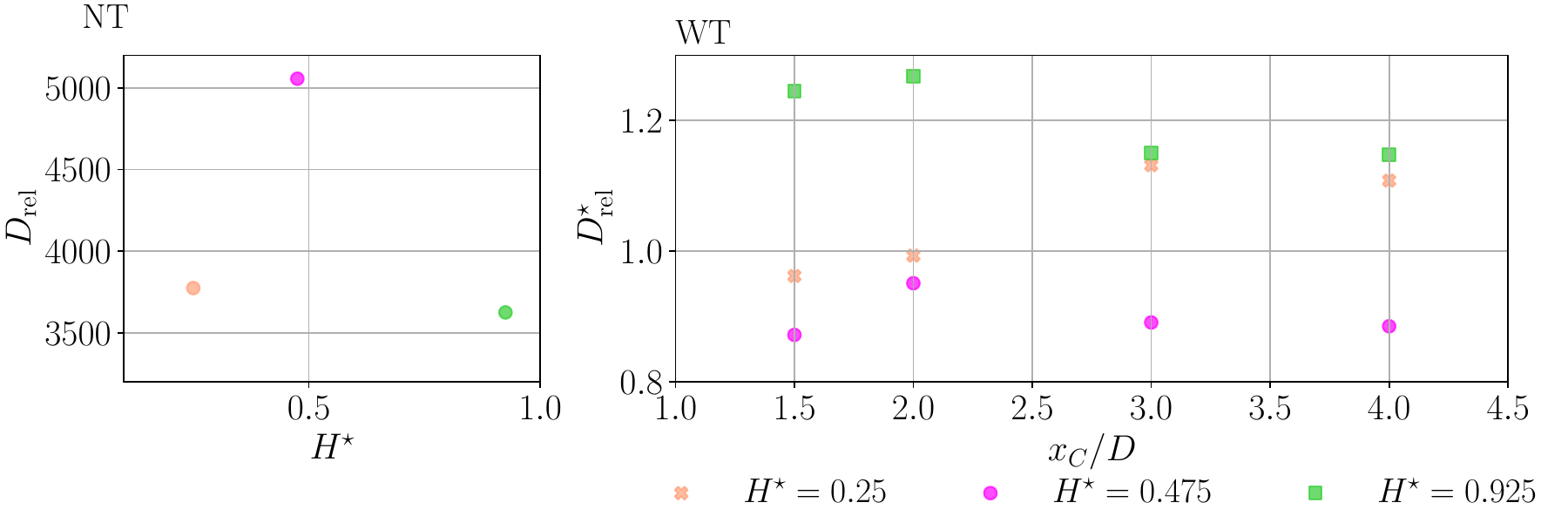}
    \caption{Left-hand pane: Cumulative damage metric $D_{rel}$, as defined in \eqref{eq:DREL}, at the location of maximum mean stress $y = L_C$. 
    Right-hand pane: $D_{rel}$ for the various cases with the wind turbine upstream of the conductor normalised by the baseline $D_{rel}$ such that $D^\star_{rel} > 1$ indicates that more damage is accumulated with the turbine present and $D^\star_{rel} < 1$ indicates that less damage is accumulated when the turbine is present.
    (a) $m =1.5$ and (b) $m=2$ where $m$ is the exponent from the definition of $D_{rel}$ in \eqref{eq:DREL}.}
    \label{fig:drel_plots}
\end{figure*}

The cumulative effect of the damage induced by the various stress cycles at location $y = L_C$ illustrated in the rainflow counting plots of figures \ref{fig:3drainflow} and \ref{fig:2drainflow} are presented, through the damage metric $D_{rel}$ defined in \eqref{eq:DREL}, in figure \ref{fig:drel_plots}.
The results are presented for two exponents $m$ in \eqref{eq:DREL} $m = $ 1.5 (a) and $m = 2$ (b).
In both cases the left-hand pane of the figure presents the damage metric for the baseline case in which there is no turbine upstream of the conductor.
We observe that the behaviour of $D_{rel}$ most closely resembles the r.m.s. strain fluctuation at $y = L_C$ presented in figure \ref{fig:rmspylon}, although the intermediate height conductor is exposed to significantly more cumulative damage than for the other two conductor heights.

The right hand panes of the figures present the damage metric computed when the turbine was placed upstream of the conductor relative to the baseline cases such that $D^\star_{rel} < 1$ means that the conductor was exposed to less damage when the turbine was present than when it was absent and $D^\star_{rel} > 1$ indicates that the conductor was exposed to more damage when the turbine was present.
Regardless of whether $m = 1.5$ or $m = 2$ there is always less cumulative damage in the conductor when a turbine is placed upstream for the intermediate conductor height.
Contrastingly, when the conductor height is close to the hub height of the turbine, meaning that it is continuously exposed to the turbine wake, more damage is always accumulated when the turbine is present (up to approximately 9\% for $m = 1.5$ and 25\% for $m = 2$).
In both scenarios, the maximum damage is accumulated when the turbine -- conductor separation is 2$D$, as it is for the intermediate conductor height.
It is reasonable to assume that for both $H^\star = 0.475$ and $H^\star = 0.925$ that at a distance of 2$D$ downstream of the turbine the conductor is fully immersed in the turbulent wind-turbine wake.
It thus seems apparent that from a fatigue damage perspective this is the most dangerous place to locate a conductor due to the nature of the wake at this location (likely the beginning of the break down of the large-scale coherent motions embedded within the wake such as the tip vortex system \citep[e.g.][]{biswas2024}).

The lowest conductor height $H^\star = 0.25$ is likely not fully immersed in the turbulent wake at this location (nor at $x_C = 1.5D$) and so we see that the conductor is exposed to less cumulative damage at these locations when the turbine is present than for the baseline.
It is not possible to explain this finding without interrogating the velocity field for these scenarios, however it may be related to the phenomenology of the layer of air squeezed between the wake and the ground illustrated in figure \ref{fig:squeeze}.
When the turbine is placed further upstream of the conductor $x_C \in \{3D, 4D\}$ the relative damage metric continues to grow for $H^\star = 0.25$ up to values around 15\% above the baseline.

Existing UK guidance states that conductors should not be built within 3 rotor diameters of a wind turbine and the one of the objectives of this work is to assess the validity of this guidance.
It is thus instructive to examine whether there is any particular change in the behaviour of the mechanical response of the conductor at $x_C = 3D$.
At the lowest conductor height there is an increase in the r.m.s. strain fluctuation in the cable when the wind turbine is positioned closer than 3$D$.
However, this does not translate into any enhanced accumulation of fatigue damage since $x_C = 3D$ is the location at which our damage metric was maximised.
For all other conductor heights the mean and r.m.s. strain fluctuation did not appreciably change from its value when the turbine is placed 3$D$ upstream of the conductor.
At the highest conductor height there was an increase in accumulated fatigue damage when the turbine was moved closer than $3D$ but for the two lower conductor heights either there was less accumulated fatigue or the increase was trivial since the accumulated damage remained lower than the baseline no-turbine case.
In this context, and based on the present experimental results, we thus conclude that there is nothing special about a turbine -- conductor spacing of 3$D$.

\section{Conclusions}
This work represents the most complete set of data yet acquired examining the issue of aerodynamic-induced fatigue of OHCs situated downstream of a wind turbine, although its limitations should be acknowledged.
These limitations include the frequency response of the measurement technique and the use of EPDM rubber for the conductor model.
The latter has significant advantages, namely that the low Young's modulus of this material enabled large strains to be measured in the conductor and hence enabled the acquisition of high quality (high signal-to-noise ratio) data.

The specific conclusions can be summarised as follows:
\begin{enumerate}
\item The mean strain in the cable is a function of the height of the conductor even in the absence of a wind turbine.
Given that the incident wind speed was fixed this difference is due to the varying turbulence intensity that the conductor is exposed to.
This has important consequences for conductors situated in different terrains since this heavily influences the nature of the atmospheric turbulence.
\item There is a location in the cable, close to the clamp, where the mean strain is approximately three times larger than the mean value in the central portion of the cable and this is the location where failure is most likely to occur (i.e. $L_c$).
\item At this critical location the mean strain in the cable was little changed from the baseline no-turbine case when a wind turbine was placed upstream of the conductor.
When the conductor was fully immersed in the turbine wake, close to the hub height, there was a noticeable decrease in the mean strain whilst at the two lowest conductor heights there was only a very minor change in the mean strain of a few percent.
\item The fluctuating strain at this critical location was very dependent on the conductor height.
When the conductor was fully immersed in the turbulent wind-turbine wake the strain fluctuation slowly increased as the turbine was placed further upstream of the conductor.
This is in keeping with recovery of the wake yet it always remained below the baseline, likely due to the reduced wind speed caused by the wake's velocity deficit.
When the turbine was closest to the conductor then the two lowest conductor heights saw small increases in the r.m.s. strain fluctuation of up to 4\% which subsequently decreased to the baseline level (intermediate conductor height) or below (lowest conductor height).
\item Spectral analysis of the strain fluctuations reveals that they are dominated by vibrations at the cable's natural frequency, or its first two harmonics, which is indicative of aeolian vibration.
The fact that the spectral peaks shifted to lower frequencies, and increased in amplitude, when the wake immersed the conductor is indicative of the fact that the aeolian vibration is enhanced by the slower moving air within a wind-turbine wake.
There was little evidence to suggest that the presence of a wind-turbine wake significantly enhances the turbulent buffeting of the conductor.
\item Rainflow counting of the bending stress cycles that the critical portion of the cable was exposed to enabled some relative fatigue analysis to be conducted.
This showed that when fully immersed in the turbine's wake the accumulated fatigue damage in the cable increased with respect to the baseline for the lowest and highest conductor heights.
When the conductor is lower to the ground this effect is less pronounced and in fact positioning the conductor closer to the turbine decelerates the accumulation of fatigue damage.
However, if the baseline is changed to a turbine -- conductor spacing of three turbine diameters, which is the current UK guidance, then it would appear that unless the conductor is to be built at a height that is similar to the hub height of the turbine then fatigue lifetimes are likely to be extended at any other distance, i.e. if the conductor is built closer or further away from the turbine.
\end{enumerate}

\vspace{1cm}
\noindent
\emph{Acknowledgments} - this work was funded by SSE Networks Transmission (SSENT) through a consultancy contract with Imperial Consultants and undertaken on an independent basis by the authors.
The authors would like to thank Glaiza Shaw at SSENT in particular for her help with setting up the project, providing the schedule of works, and assisting during the acquisition of the data.
The authors gratefully acknowledge the expertise and advice offered by several colleagues in the Department of Aeronautics at Imperial College London, particularly Paul Howard, Ricardo Huerta Cruz and Will McArdle, during wind tunnel model setup, and Dominica Rohozinski for helping with the design and manufacturing of the pylon models.
OB would also like to acknowledge funding from the Engineering and Physical Sciences Research Council (EPSRC) through grant no. EP/V006436/1.
Use was also made of EPSRC grant no. EP/L024888/1 for access to the National Wind Tunnel Facility (NWTF).

\bibliographystyle{plainnat}
\bibliography{bibliography}
\end{document}